\documentclass[useAMS,usenatbib]{mn2e}
\usepackage{amssymb}
\usepackage{deluxetable}

\title[VLBA observations of HFP radio sources]{Physical properties in
  young radio sources. VLBA observations of high frequency peaking
  radio sources}
\author[M. Orienti \& D. Dallacasa ]
  {M. Orienti$^{1,2}$\thanks{E-mail: orienti@ira.inaf.it},
D. Dallacasa$^{2,1}$\\
$^1$INAF -- Istituto di Radioastronomia, via Gobetti 101, I-40129, Bologna,
Italy \\
$^2$Dipartimento di Fisica ed Astronomia, Universit\`a di Bologna, via Ranzani 1,
I-40127, Bologna, Italy \\
}
\date{Received \today; accepted ?}

\pagerange{\pageref{firstpage}--\pageref{lastpage}} \pubyear{2002}

\def\LaTeX{L\kern-.36em\raise.3ex\hbox{a}\kern-.15em
    T\kern-.1667em\lower.7ex\hbox{E}\kern-.125emX}

\begin{document}

\label{firstpage}

\maketitle

\begin{abstract}
Multifrequency Very Long Baseline Array (VLBA) observations were
performed to study the radio morphology and the synchrotron spectra of
four high frequency peaking radio sources. They are resolved in
several compact components and the radio emission is dominated by the
hotspots/lobes. The core region is unambiguously detected in 
J1335+5844 and J1735+5049. The spectra of the main source components peak
above 3 GHz. 
Assuming that the spectral peak
is produced by synchrotron self-absorption, we estimate 
the magnetic field directly from observable quantities and in half of
the components it agrees with the
equipartition field, while in the others
the difference exceeds an order of magnitude.
By comparing the
physical properties of the targets with those of
larger objects we found that the luminosity increases with
the linear size for sources smaller than a few
kpc, while it decreases for larger objects. 
The asymmetric sources J1335+5844 and
J1735+5049 suggest that
the ambient medium is inhomogeneous and is 
able to influence the evolution of the radio emission even during its
first stages. The core luminosity increases with
the linear size for sources up to a few kpc, while it seems constant
for larger sources suggesting an evolution independent from
the source total luminosity. 

\end{abstract}


\section{Introduction}

Understanding how radio emission originates and evolves 
in extragalactic radio sources is one of the greatest challenges 
faced by modern astrophysics.\\
In the evolutionary scenario, the size of a radio source is strictly 
related to its age. 
Given their intrinsically compact sizes, usually smaller than the host
galaxy, 
the population of compact
steep spectrum (CSS) and gigahertz-peaked spectrum (GPS) radio sources
were proposed as the progenitors of the classical radio galaxies
\citep{fanti95,readhead96,snellen00b}. Indeed their radio
structure is usually a scaled-down version of the Fanaroff-Riley
type-II galaxy \citep{fanaroff74}. It is mostly accepted that, at
least the two-sided 
(or symmetric) objects of this class, represent the young stage in the
radio source evolution. Depending on their linear size they are termed
as compact symmetric objects (CSO) or medium symmetric objects (MSO)
if they are smaller or larger than 1 kpc, respectively
\citep{fanti95}, although for a strict application of the definition
the core must be detected.\\
The genuine youth of these
objects is strongly supported by the determination of the kinematic
and radiative ages in about twenty of the most compact CSS/GPS
objects, which were found to be a few thousand years or less
\citep[e.g.][]{owsianik98,polatidis03,giroletti09,murgia03}. \\
The determination of the physical properties in objects which are 
at the beginning of their evolution is crucial for setting tight
constraints on the initial conditions of the radio emission. To this
aim the selection of radio sources in the very first stages of the
radio emission is necessary. Following the empirical
anti-correlation between the spectral peak frequency and the source
size (i.e. the source age) found by \citet{odea97}, 
the youngest objects should be sought
among high frequency peaking (HFP) radio sources. A sample of young HFP
candidates was selected by \citet{dd00} on the basis of their inverted
spectrum between 1.4 and 5 GHz. The contamination from variable,
boosted objects was removed on the basis of their
variability, morphology and polarization properties
\citep{tinti05,mo06,mo07,mo08a}. The sample of genuine young HFP 
candidates is reported in \citet{mo08a}. 
A crucial requirement for unambiguously classifying a source as a
genuinely young radio source is the detection of the
core component.
Among the sources in the sample of young HFP candidates \citep{mo08a},
the core has been detected in three objects: 
J0111+3906, J0650+6001, and J1511+0518. 
Studies of the proper motion
in J0111+3906 \citep{owsianik98}, J0650+6001 \citep{modd10}, and in
J1511+0518 \citep{mo08,an12} estimate a kinematic age of a few
hundred years, indicating that HFPs are at an early stage of their radio
evolution/growth.\\
In this paper we present multifrequency VLBA observations of
an additional four HFP objects with the aim of detecting the core
component and determining the physical properties, such as the peak
frequency, the magnetic field and the luminosity, which are important
for constraining the evolutionary models.\\
The paper is organized as follows: Section 2 describes the
observations and the data reduction; Section 3 presents the
description of the target sources and the results from the spectral
analysis; discussion and summary are presented in Sections 4 and 5. An
appendix with Very Large Array (VLA) flux densities of sources from
the bright HFP sample is provided.\\
 
Throughout this paper, we assume 
$H_{0} = 71$ km s$^{-1}$ Mpc$^{-1}$, $\Omega_{\rm M} = 0.27$,
$\Omega_{\Lambda} = 0.73$, in a flat Universe. The spectral index is
defined as $S$($\nu$) $\propto \nu^{- \alpha}$. \\ 

\section{Observations and Data Reduction}

VLBA observations of the four HFP sources J0037+0808, J1335+5844,
J1735+5049, and J2203+1007 from the bright HFP sample \citep{dd00}
were performed in full polarization with a recording band width of 32 MHz at 256
Mbps in L band (1.4 and 1.6 GHz), S band (2.2 GHz), C band (4.6
and 5.0), X band (8.1 and 8.4 GHz), and U band (15 GHz). The sources
J1335+5844 and J1735+5049 were bright enough to allow detection at
high frequency, and observations at 22 GHz (K band) were
performed for these two sources. \\
Observations
were carried out between January 2010 and January 2011 in four
different runs (one per source), for a 
total observing time of 36 hours. In each observing run, the target
source was observed interleaved with its calibrator, in order to
optimize the uv-coverage, and cycling through
frequencies. This allows us to consider the observations
at the various frequencies almost simultaneous.
In L, C, and X bands,
the intermediate frequencies are separated by about 300 MHz in order
to improve the frequency coverage of the source spectra. \\
The correlation was performed at the VLBA correlator in Socorro and
the data reduction was carried out with the NRAO's Astronomical Image
Processing System (\texttt{AIPS}) package.  
After the application of system temperature and antenna gain
information, the amplitudes were checked by using the data of the
fringe finders DA\,195 (J0555+3948), 4C\,39.25 (J0927+3902), 3C\,345
(J1642+3948), and 3C\,454.3 (J2253+1608). All the gain corrections 
were found to be within 5\% in L, C, and X bands, 7\% in S and U
bands, and 10\% in K band, 
which can be conservatively assumed as the absolute calibration error
($\sigma_{\rm c}$).
The sources were strong enough to allow fringe fitting.
Final images were produced after a number of phase self-calibration
iterations. At the end of the self-calibration iteration we applied the
amplitude calibration to remove residual
systematic errors. \\
The final rms noises (1$\sigma$) are
generally between 0.08 and 0.5 mJy/beam, being worst at the highest
frequency.

\begin{table*}
\caption{VLBA total flux density. Column 1: source name; Col. 2:
  redshift. p=photometric redshift; Cols. 3, 4, 5, 6, 7, 8, 9, 10, and
11: VLBA flux density at 1.4, 1.6, 2.2, 4.6, 5.0, 8.1, 8.4, 15, and 22
GHz, respectively; Col. 12: reference for the redshift. 1=Sloan
Digital Sky Survey Data Release 9 \citep[SDSS DR9;][]{ahn12};
2=\citet{britzen08}; 3=\citet{healey08}.}
\begin{center} 
\begin{tabular}{cccccccccccc}
\hline
Source& z &$S_{\rm 1.4}$&$S_{\rm 1.6}$&$S_{\rm 2.2}$&$S_{\rm
  4.6}$&$S_{\rm 5.0}$&$S_{\rm 8.1}$&$S_{8.4}$&$S_{\rm 15}$&$S_{\rm
  22}$& Ref.\\
  & &mJy&mJy&mJy&mJy&mJy&mJy&mJy&mJy&mJy& \\
 (1)&(2)&(3)&(4)&(5)&(6)&(7)&(8)&(9)&(10)&(11)&(12)\\
\hline
&&&&&&&&&&&\\
J0037+0808&  - &94$\pm$5& 111$\pm$6& 177$\pm$9& 264$\pm$13&
270$\pm$13& 260$\pm$13& 246$\pm$12& 158$\pm$11& - &  -\\
J1335+5844& 0.58p& 280$\pm$14& 349$\pm$17& 578$\pm$40& 653$\pm$33&
659$\pm$33& 638$\pm$33& 624$\pm$31& 498$\pm$35& 318$\pm$32& 1\\
J1735+5049& 0.835& 457$\pm$23&471$\pm$23& 573$\pm$29& 873$\pm$44&
890$\pm$44& 906$\pm$45& 921$\pm$46& 703$\pm$49& 637$\pm$64& 2\\
J2203+1007& 1.005& 115$\pm$6& 133$\pm$7& 213$\pm$15& 277$\pm$14&
282$\pm$14&201$\pm$10& 200$\pm$10& 108$\pm$8& - & 3\\
&&&&&&&&&&&\\
\hline
\end{tabular}
\end{center}
\label{vlba-tot} 
\end{table*}

\begin{table*}
\caption{Observational parameters of the source components. Column 1:
  source name; Col. 2: source component; Cols. 3, 4, 5, 6, 7, 8, 9, 10, and
  11: VLBA flux density at 1.4, 1.6, 2.2, 4.6, 5.0, 8.1, 8.4, 15, and
  22 GHz, respectively; Cols. 12 and 13: major and minor axis,
  respectively; Col. 14: position angle of the major axis.  }
{\scriptsize
\begin{center}
\begin{tabular}{cccccccccccccc}
\hline
Source&Comp.&$S_{\rm 1.4}$&$S_{\rm 1.6}$&$S_{\rm 2.2}$&$S_{\rm
  4.6}$&$S_{\rm 5.0}$&$S_{\rm 8.1}$&$S_{8.4}$&$S_{\rm 15}$&$S_{\rm
  22}$&$\theta_{\rm maj}$&$\theta_{\rm min}$&p.a.\\
 & &mJy&mJy&mJy&mJy&mJy&mJy&mJy&mJy&mJy&mas&mas $^{\circ}$\\
(1)&(2)&(3)&(4)&(5)&(6)&(7)&(8)&(9)&(10)&(11)&(12)&(13)&(14)\\
\hline
&&&&&&&&&&&&&\\
J0037+0808&E\tablenotemark{a}&  94$\pm$5& 111$\pm$6& 177$\pm$9& 223$\pm$11& 223$\pm$11& 231$\pm$11& 209$\pm$10&
148$\pm$10& - &0.266$^{+0.001}_{-0.001}$&0.100$^{+0.007}_{-0.007}$&58$\pm$1\\
          &E1& - &  - & -  & -  & -  &  - &  166$\pm$8&  105$\pm$7& -& 0.577$^{+0.003}_{-0.004}$&0.206$^{+0.012}_{-0.012}$&52$\pm$1\\
          &E2& - &  - & - & - & - &  - &50$\pm$3  &43$\pm$3 & - & -& - & - \\
          & W& - &  - &  - &  38$\pm$2&  47$\pm$2&  39$\pm$1&  37$\pm$2& 10$\pm$1 & - & - & - \\
J1335+5844&N\tablenotemark{a} &  91$\pm$4& 108$\pm$5& 171$\pm$12& 306$\pm$15& 330$\pm$16& 459$\pm$23& 457$\pm$23& 447$\pm$31& 304$\pm$30&4.2&1.6&-158\\
          &N1&  - &  - &  - & 198$\pm$10& 262$\pm$13& 423$\pm$21& 423$\pm$21& 372$\pm$26&
271$\pm$27&0.296$^{+0.001}_{-0.001}$&0.136$^{+0.001}_{-0.001}$&142$\pm$1\\
          &N1b&  - &  - &  - &  - &  - &  - &  - &  60$\pm$4&  23$\pm$2& - & - & -\\
          &N2&   - &  - &  - & 108$\pm$5&  68$\pm$3&  34$\pm$2&  34$\pm$2&  15$\pm$1&
9$\pm$1&0.307$^{+0.017}_{-0.016}$&0.251$^{+0.009}_{-0.012}$&163$\pm$14\\ 
          &C &   - &  - &  - &  - &  - &  - & 1.3$\pm$0.2& 1.6$\pm$0.2& 1.6$\pm$0.2& - & - & -\\
          &S &  189$\pm$9& 241$\pm$12& 407$\pm$28& 347$\pm$17& 329$\pm$16& 179$\pm$9& 166$\pm$8&  49$\pm$3&  14$\pm$2&1.205$^{+0.008}_{-0.008}$&0.946$^{+0.009}_{-0.010}$&114$\pm$2\\
J1735+5049&N &   - & - &  64$\pm$4& 275$\pm$14& 280$\pm$14& 220$\pm$11& 212$\pm$11&  94$\pm$7&  55$\pm$6&0.648$^{+0.012}_{-0.013}$&0.425$^{+0.016}_{-0.017}$&46$\pm$3\\
          &C &   - & - & -  &  - &  - &  - &  - &  12$\pm$1&  14$\pm$1& - & - & -\\
          &S &  457$\pm$23&471$\pm$23& 509$\pm$36& 598$\pm$30& 610$\pm$30& 686$\pm$34& 715$\pm$36& 609$\pm$43&
568$\pm$57&0.358$^{+0.002}_{-0.002}$&0.112$^{+0.002}_{-0.003}$&18$\pm$1\\
J2203+1007&E\tablenotemark{a}  &  84$\pm$4&  91$\pm$4& 130$\pm$9& 182$\pm$9& 191$\pm$9&151$\pm$8& 159$\pm$8&  90$\pm$6& -  &4.7&4.1&-75\\
          &E1 &   -&  - &  - & 148$\pm$7& 156$\pm$8& 73$\pm$4&  76$\pm$4&  56$\pm$4& -  &0.408$^{+0.006}_{-0.006}$&0.317$^{+0.011}_{-0.013}$&112$\pm$4\\
          &E2 &  - &  - &  - &  34$\pm$2&  35$\pm$2& 62$\pm$3&  69$\pm$3&  33$\pm$2& -  &1.409$^{+0.010}_{-0.011}$&0.199$^{+0.048}_{-0.063}$&88$\pm$1\\
          &E3 &  - &  - &  - &  - &  - & 15$\pm$1&   13$\pm$1&  1.0$\pm$0.2& -  & -& - & -\\
          &W\tablenotemark{a}  &  31$\pm$2&  42$\pm$2&  83$\pm$6&  95$\pm$5&  91$\pm$5& 48$\pm$2&  50$\pm$2&
18$\pm$1& - &2.9&2.9& -\\ 
          &W1 &  - &  - &  - &  - & -  & 32$\pm$2&  34$\pm$2&  11$\pm$1& - &0.757$^{+0.026}_{-0.027}$&0.369$^{+0.011}_{-0.011}$&175$\pm$2\\ 
          &W2 &  - &  - &  - &  - &  - & 16$\pm$1&  16$\pm$1&   7$\pm$1& - &0.748$^{+0.027}_{-0.021}$&0.537$^{+0.057}_{-0.071}$&87$\pm$11\\     
&&&&&&&&&&&&&\\
\hline
\end{tabular}
\end{center}}
\tablenotetext{a}{The flux density refers to the whole component. The
  angular size is measured from the lowest contours on the image plane
  and corresponds to
  1.8 times the size of a conventional Gaussian covering a similar
  area \citep{read94}. }
\label{table-vlba}
\end{table*}

\begin{figure*}
\begin{center}
\includegraphics{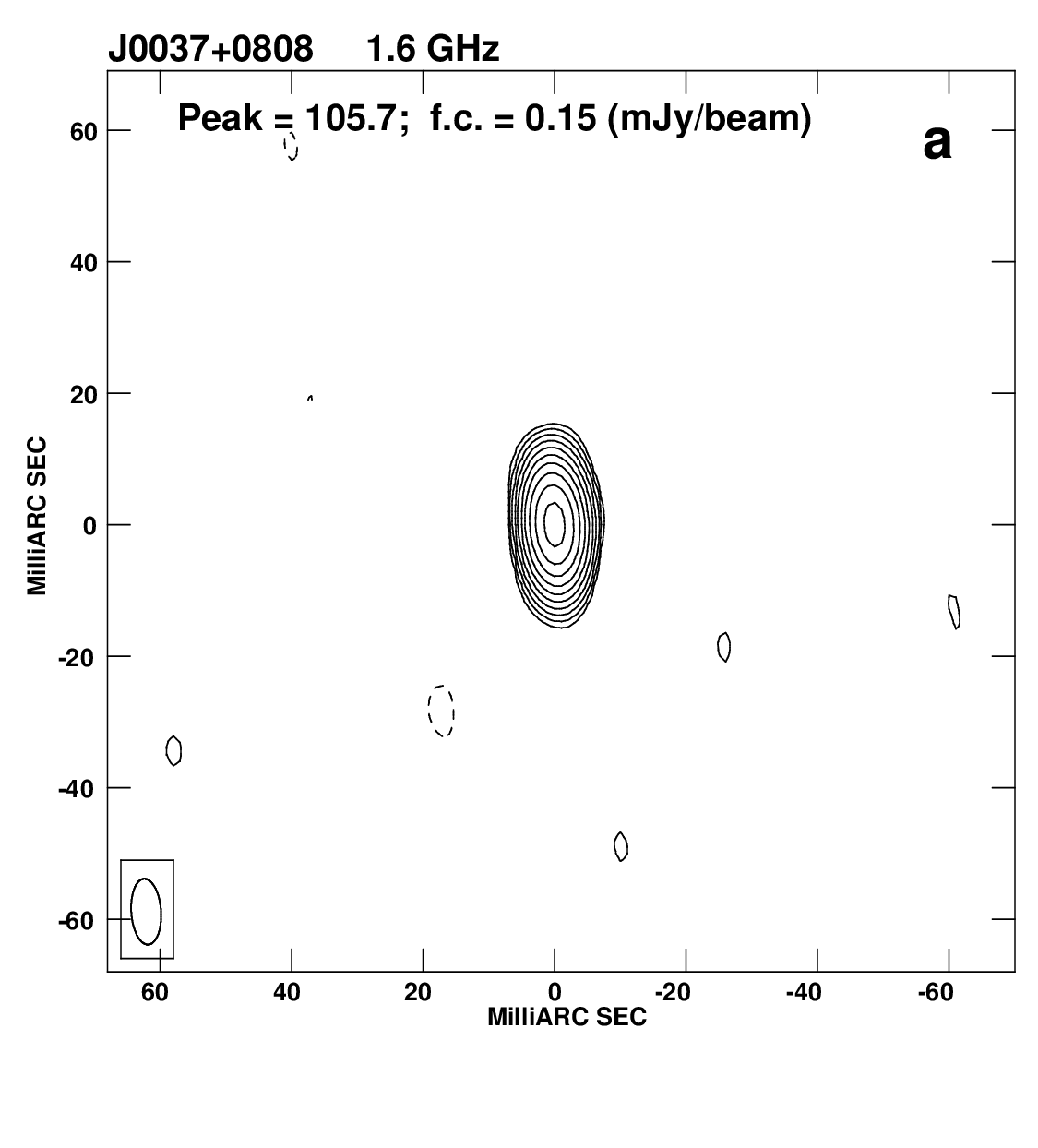}
\includegraphics{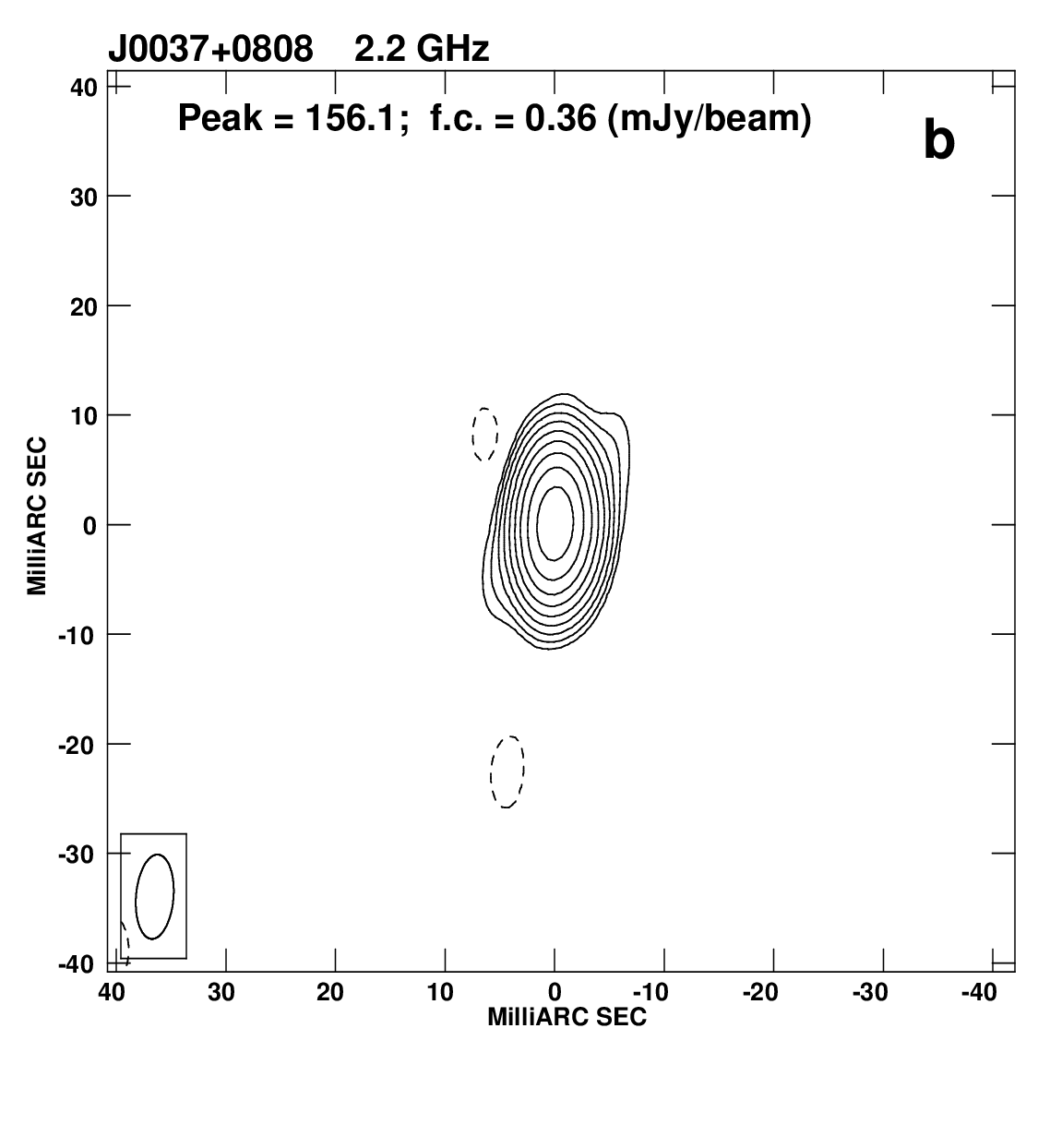}
\includegraphics{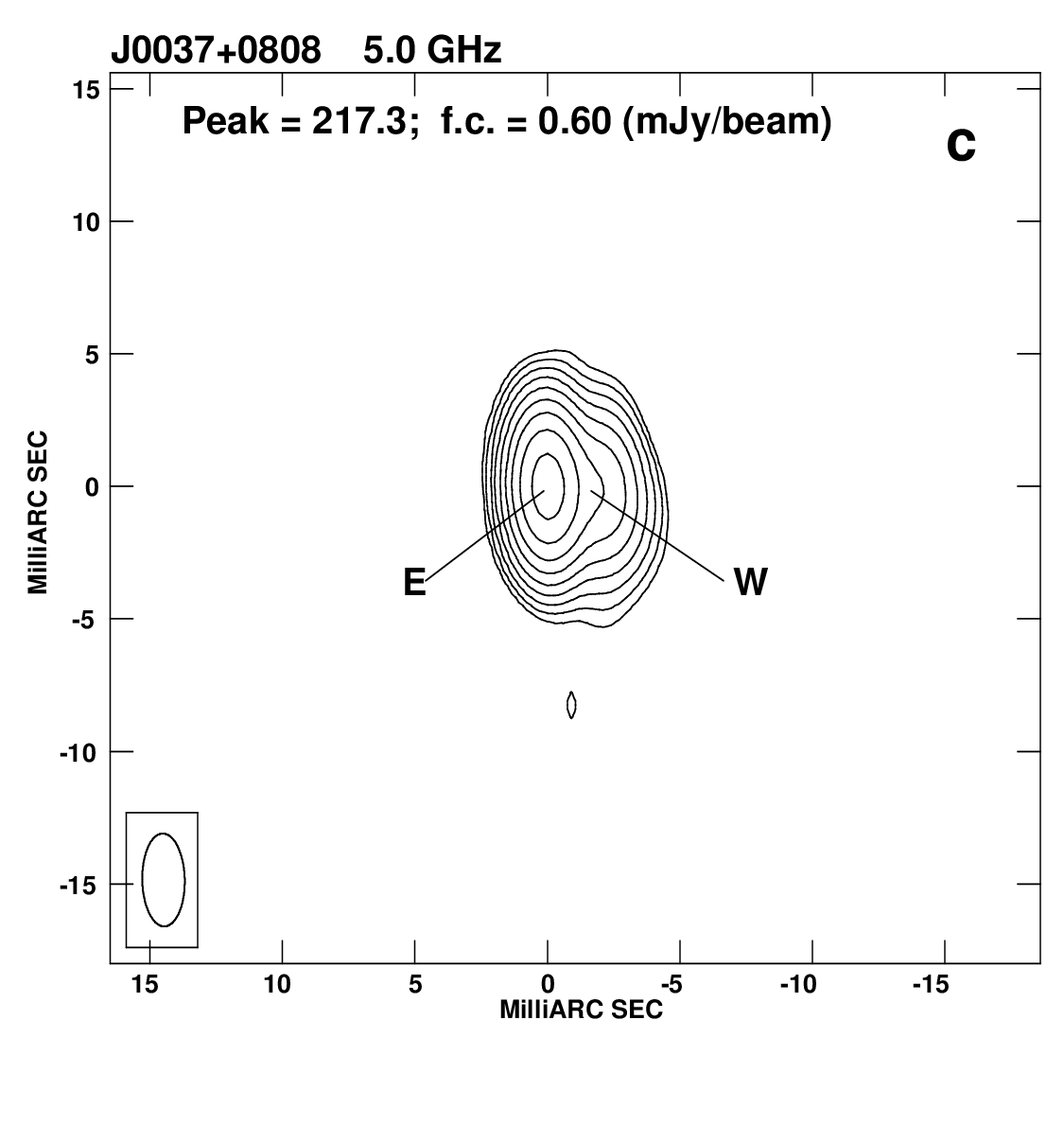}
\includegraphics{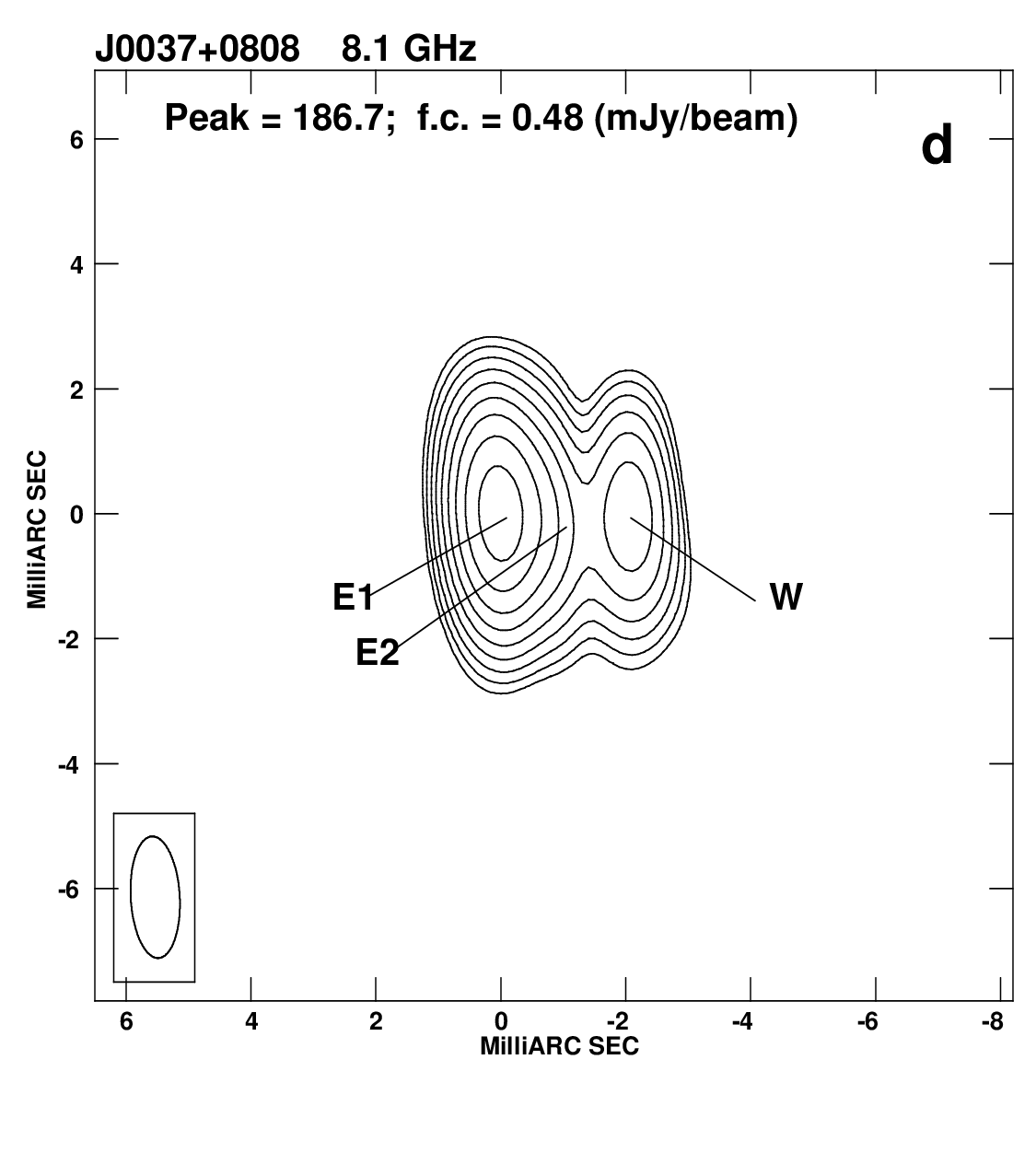}
\includegraphics{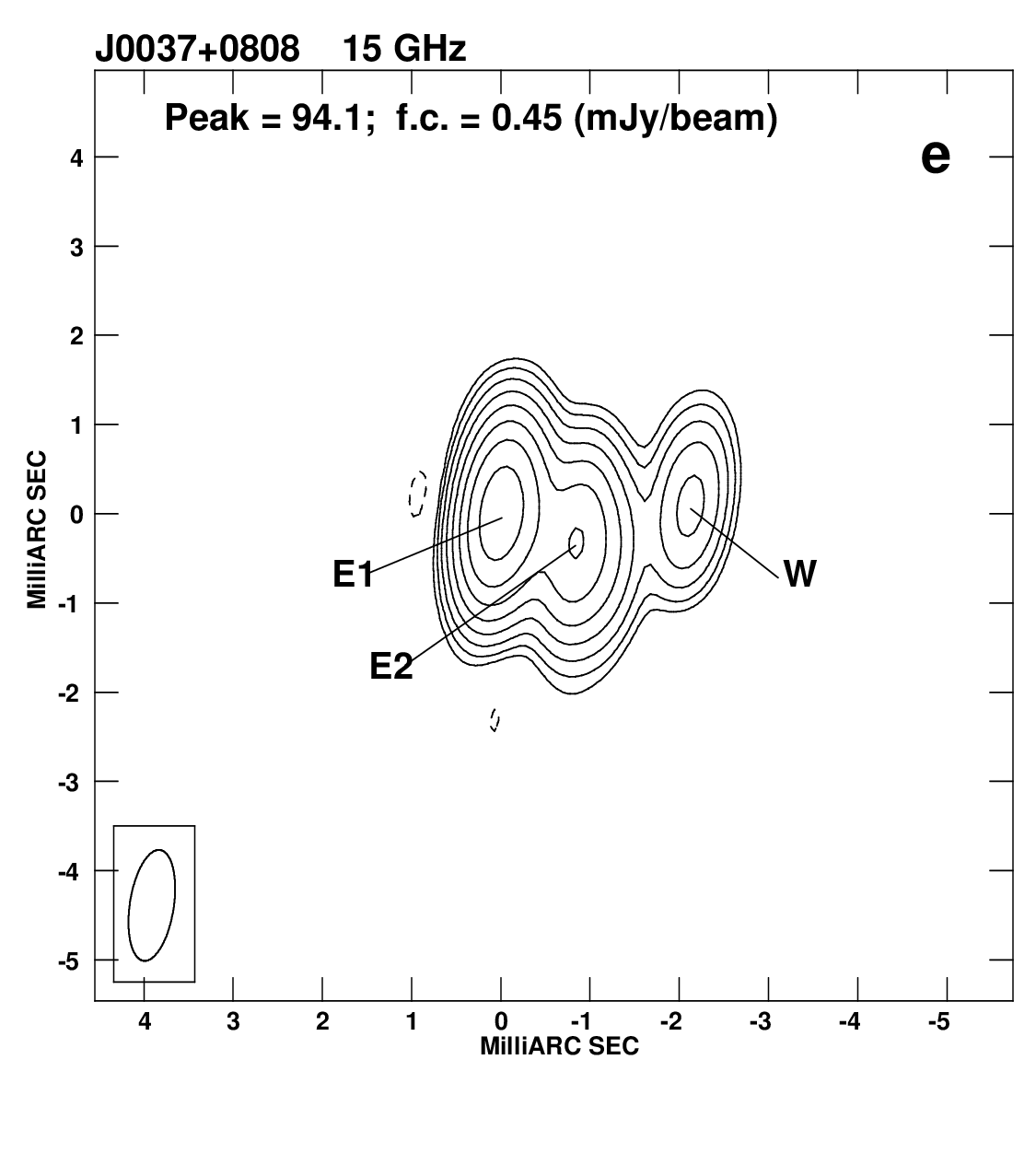}
\vspace{12.5cm}
\caption{VLBA observations of J0037+0808 at 1.6 GHz ({\bf a}), 2.2 GHz
  ({\bf b}), 5.0 GHz ({\bf c}), 8.1 GHz ({\bf d}), and 15 GHz ({\bf
  e}). On each image we provide the peak flux density in mJy/beam and the first
contour intensity (f.c.) in mJy/beam, which corresponds to three times
the off-source noise level measured on the image plane. Contours
increase by a factor of 2. The restoring beam is plotted in the bottom
left corner of each panel.}
\label{0037}
\end{center}
\end{figure*}

\begin{figure*}
\begin{center}
\includegraphics{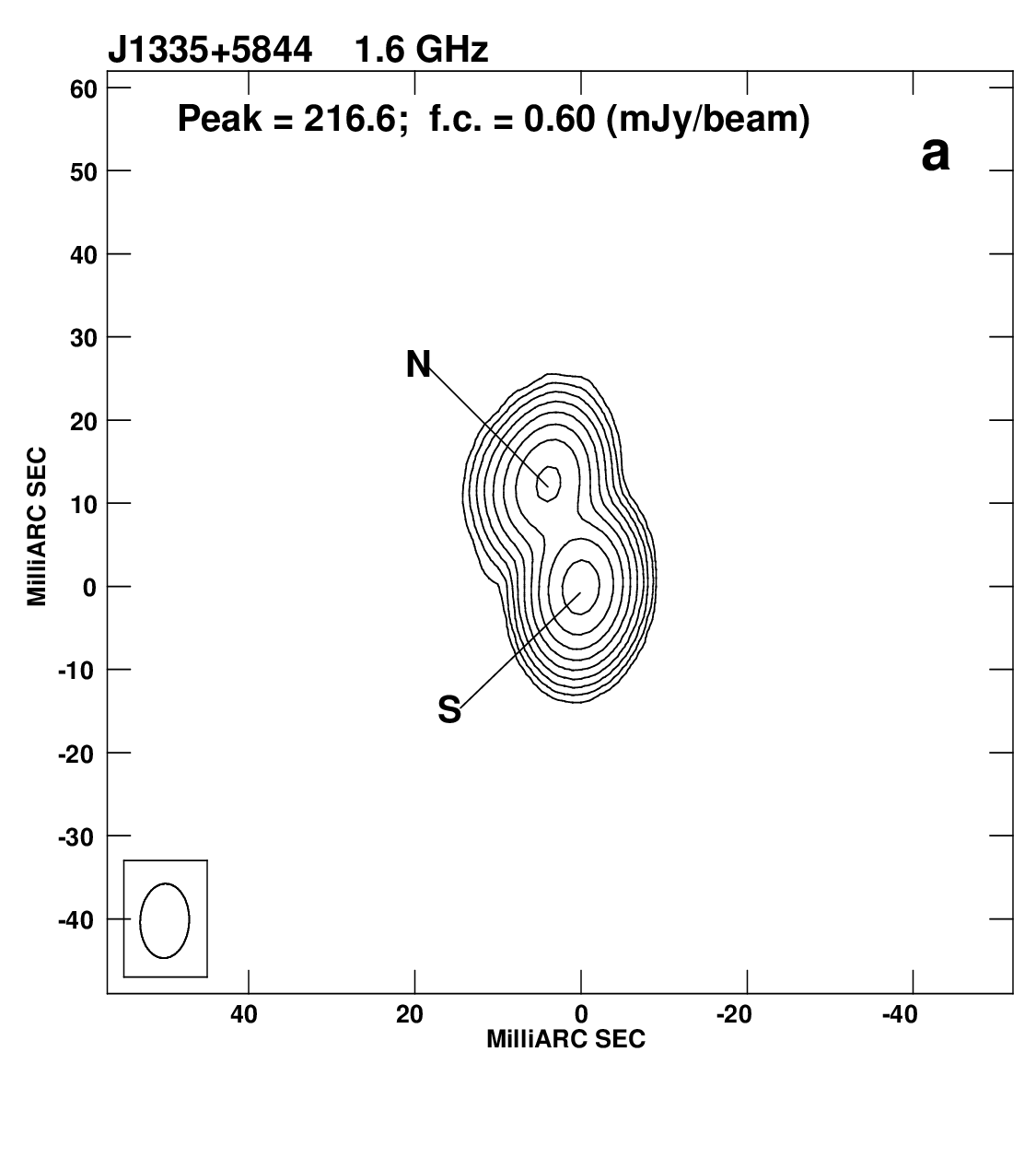}
\includegraphics{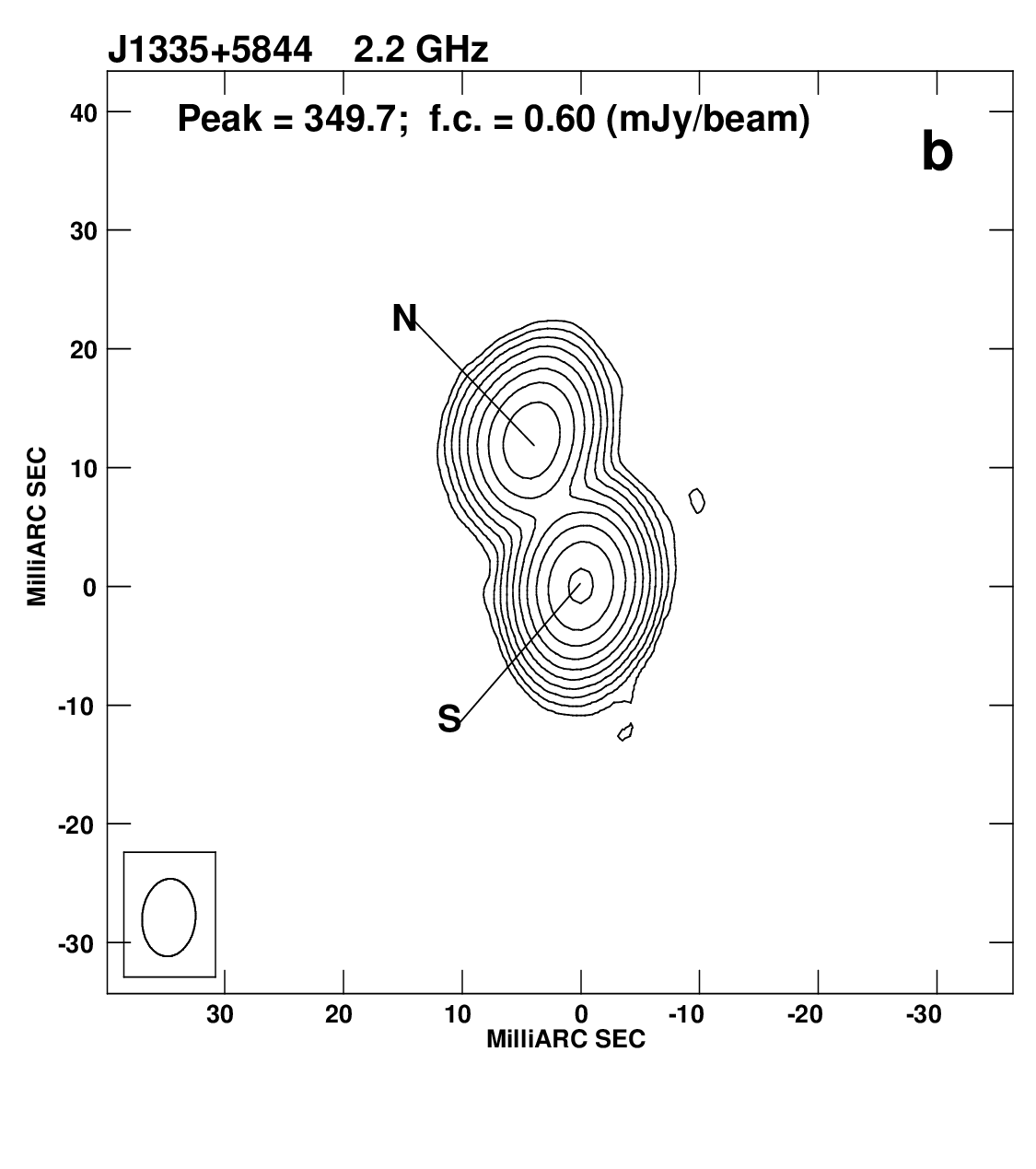}
\includegraphics{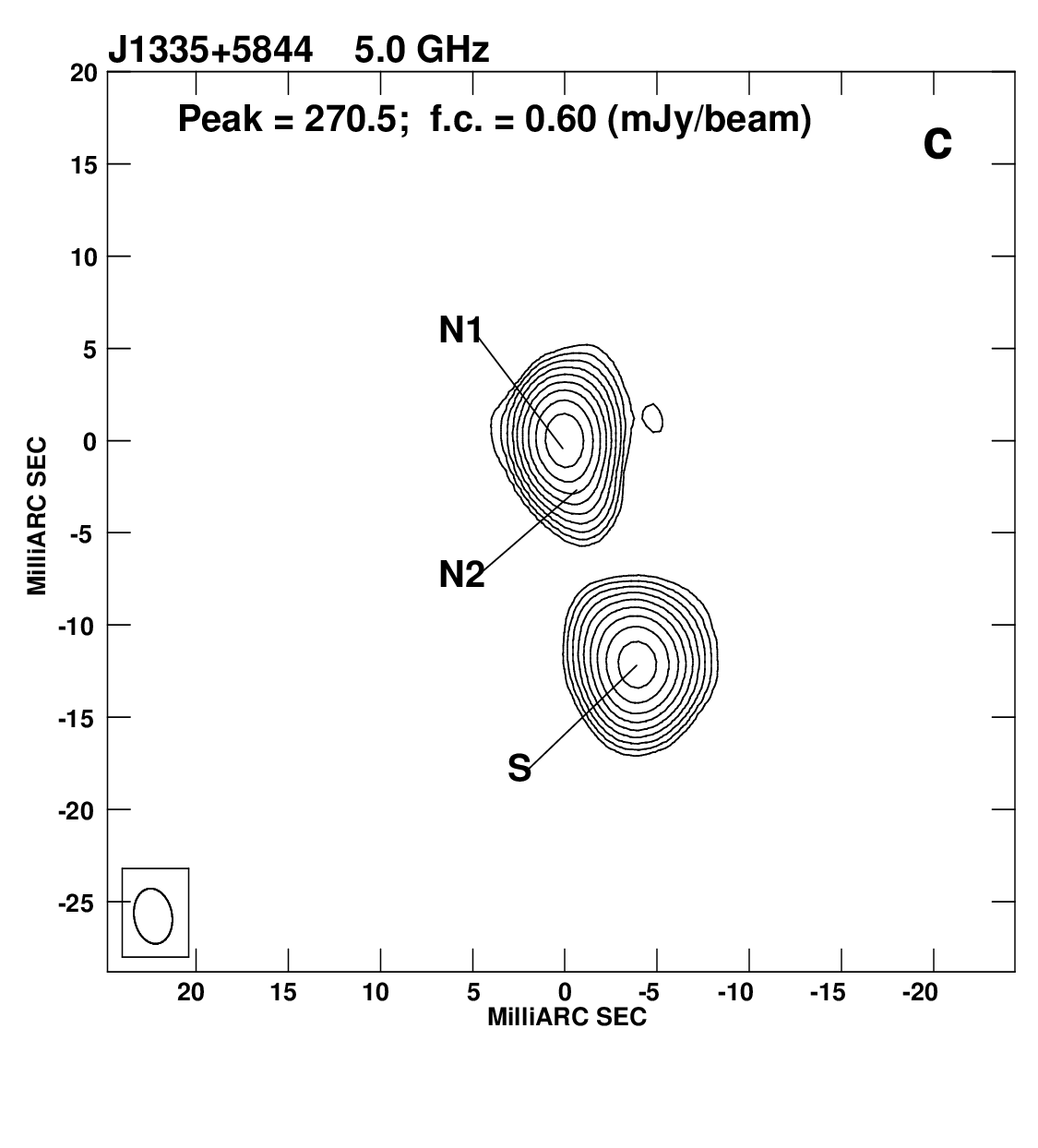}
\includegraphics{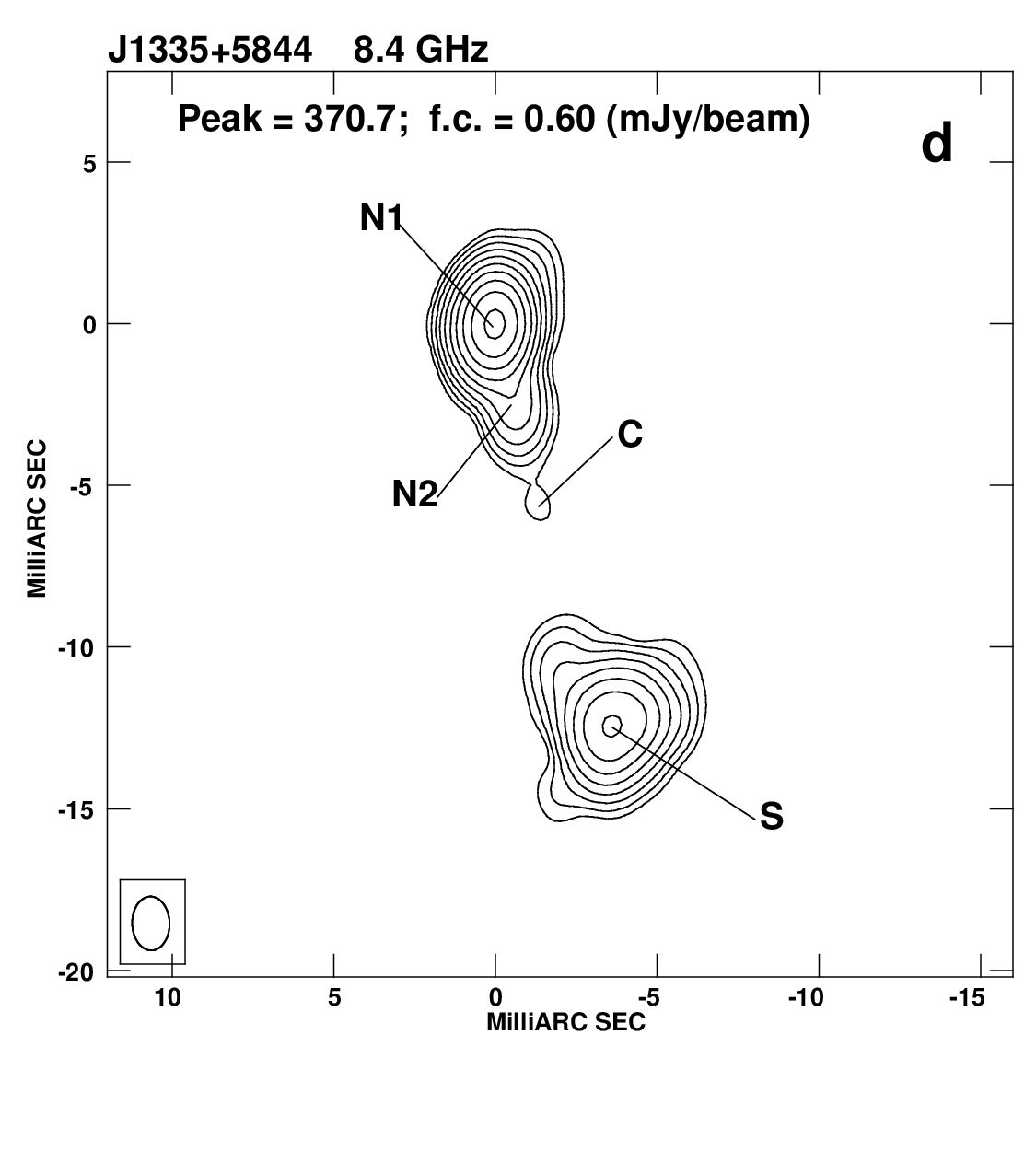}
\includegraphics{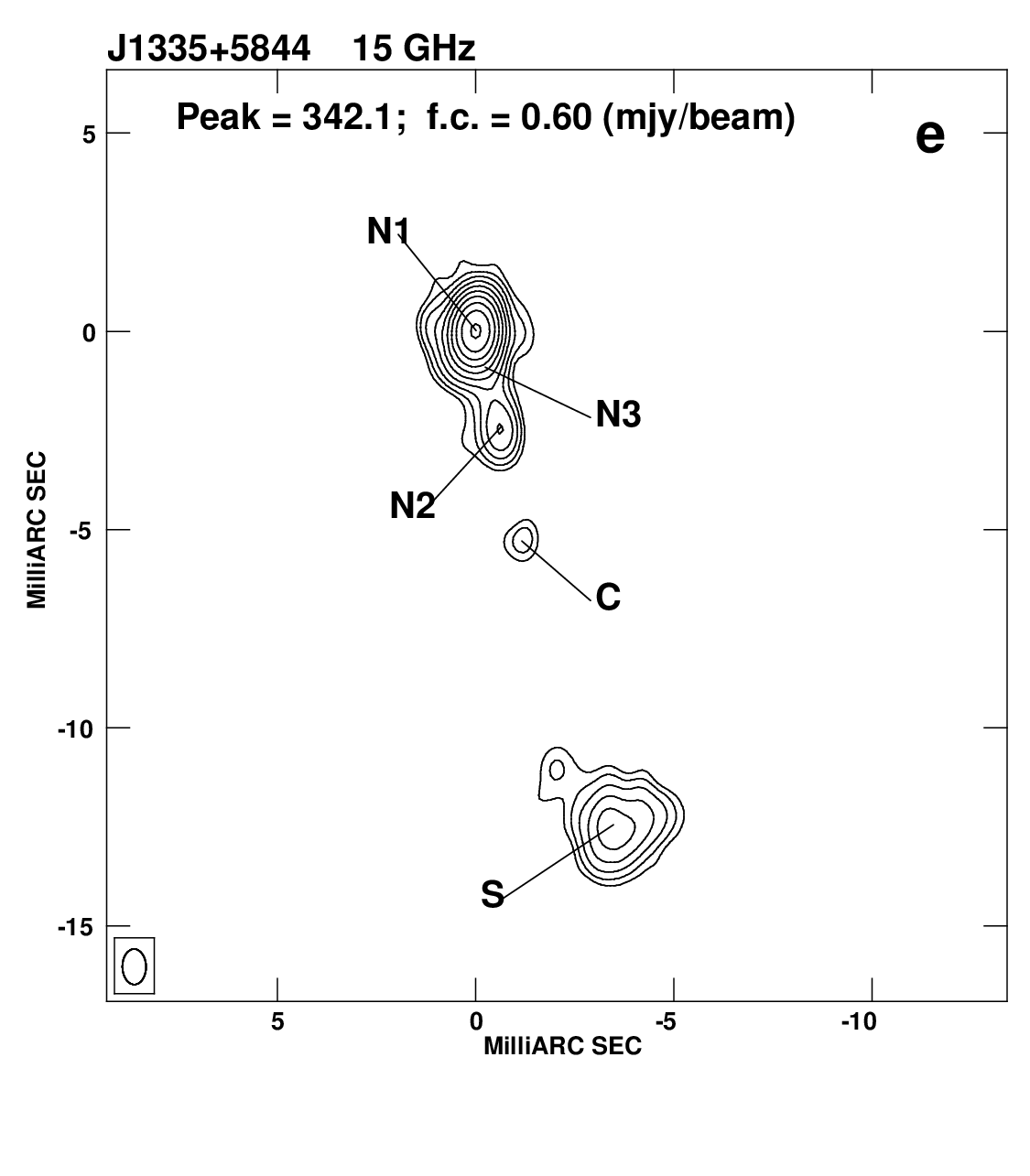}
\includegraphics{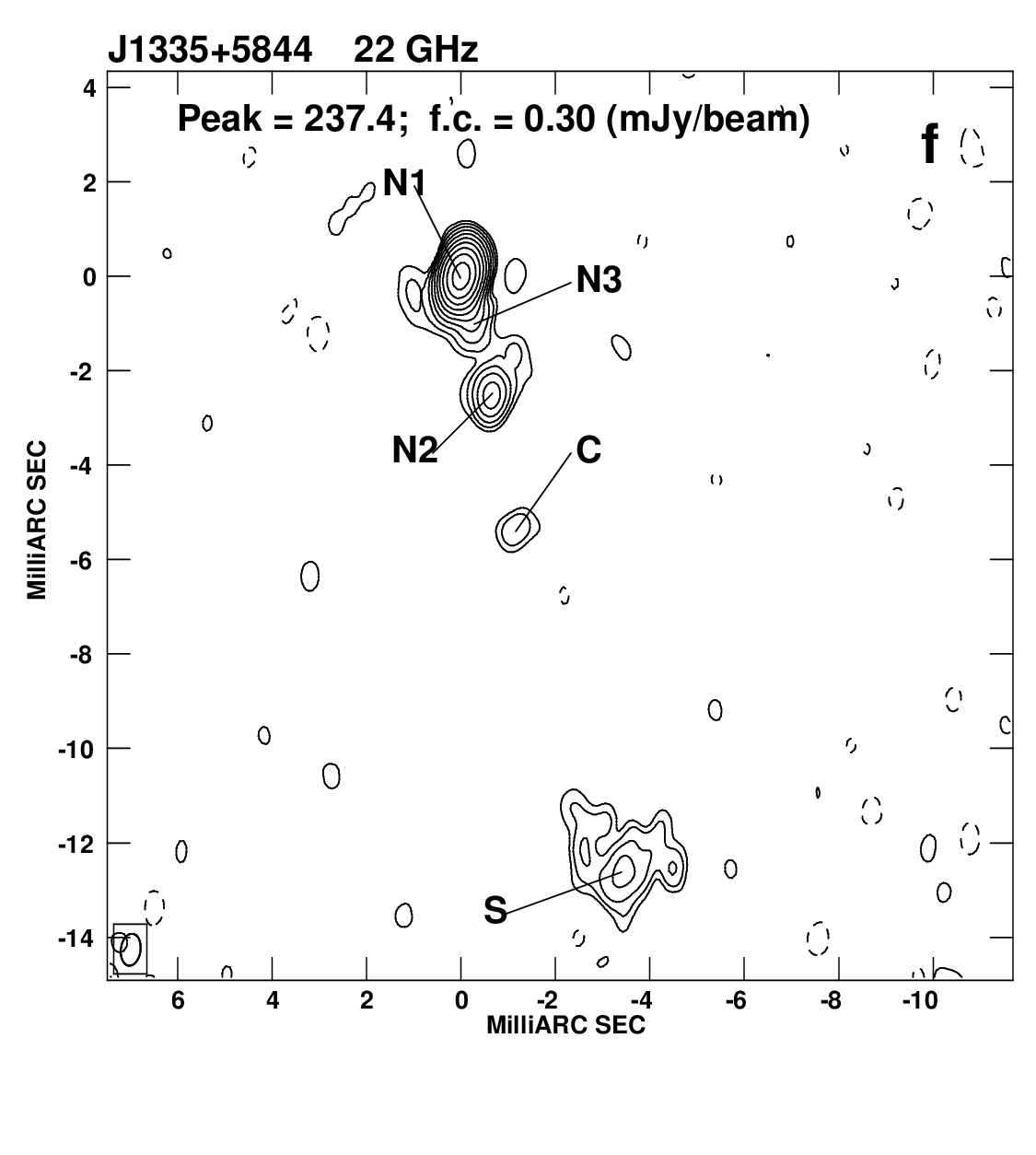}
\vspace{12.5cm}
\caption{VLBA observations of J1335+5844 at 1.6 GHz ({\bf a}), 2.2 GHz
  ({\bf b}), 5.0 GHz ({\bf c}), 8.4 GHz ({\bf d}), 15 GHz ({\bf
  e}), and 22 GHz ({\bf f}). 
On each image we provide the peak flux density in mJy/beam and the first
contour intensity (f.c.) in mJy/beam, which corresponds to three times
the off-source noise level measured on the image plane. Contours
increase by a factor of 2. The restoring beam is plotted in the bottom
left corner of each panel.}
\label{1335}
\end{center}
\end{figure*}

\begin{figure*}
\begin{center}
\includegraphics{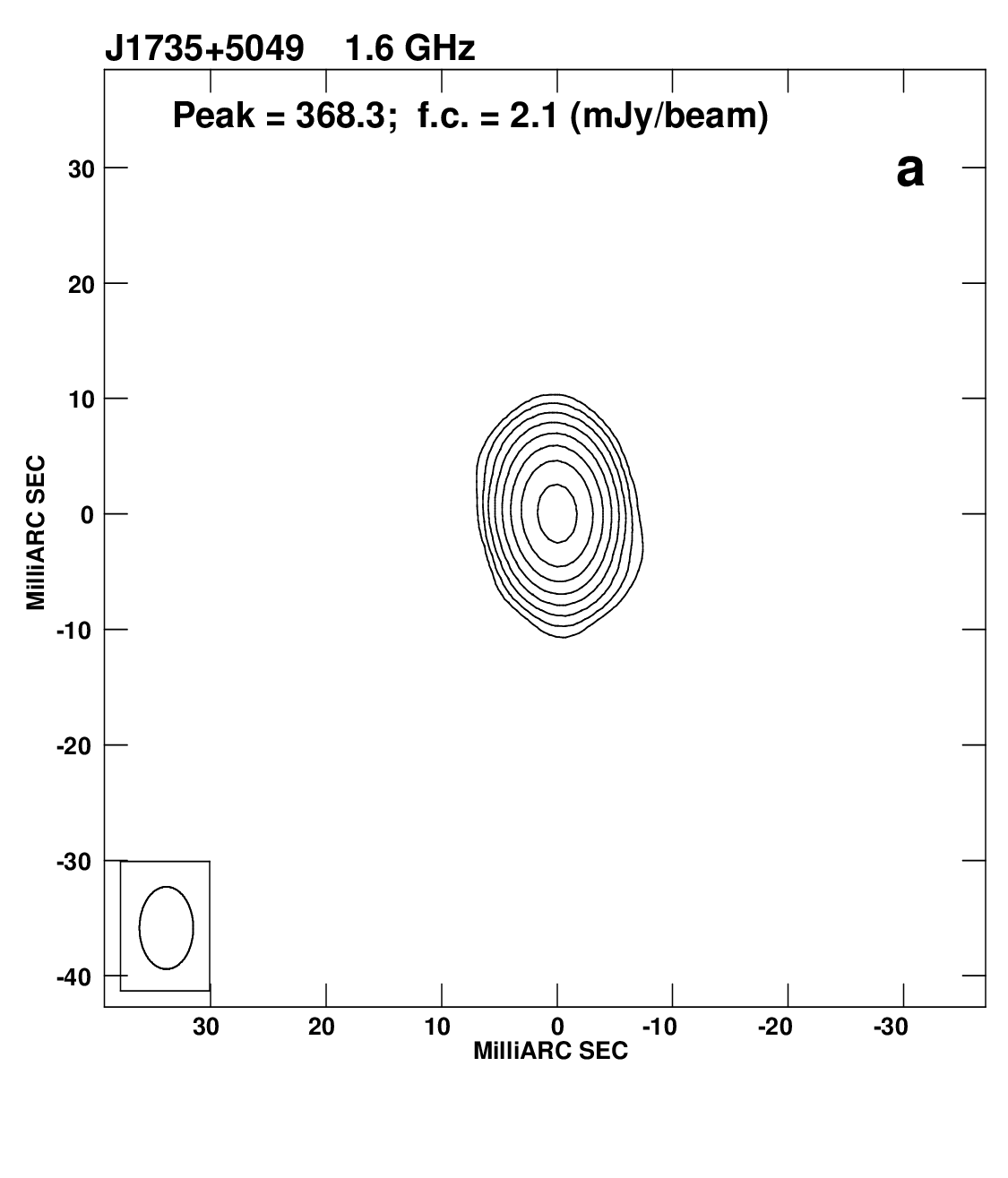}
\includegraphics{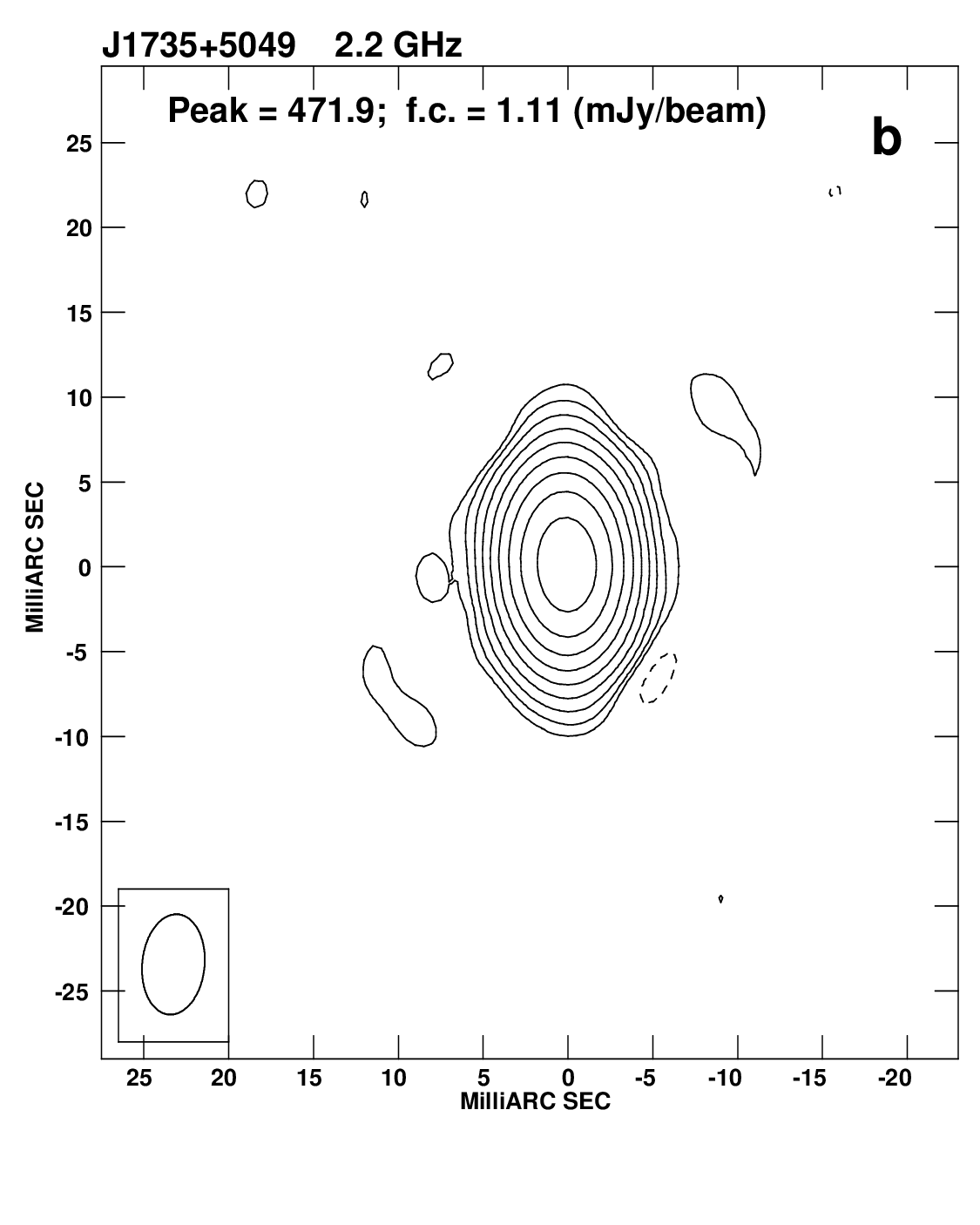}
\includegraphics{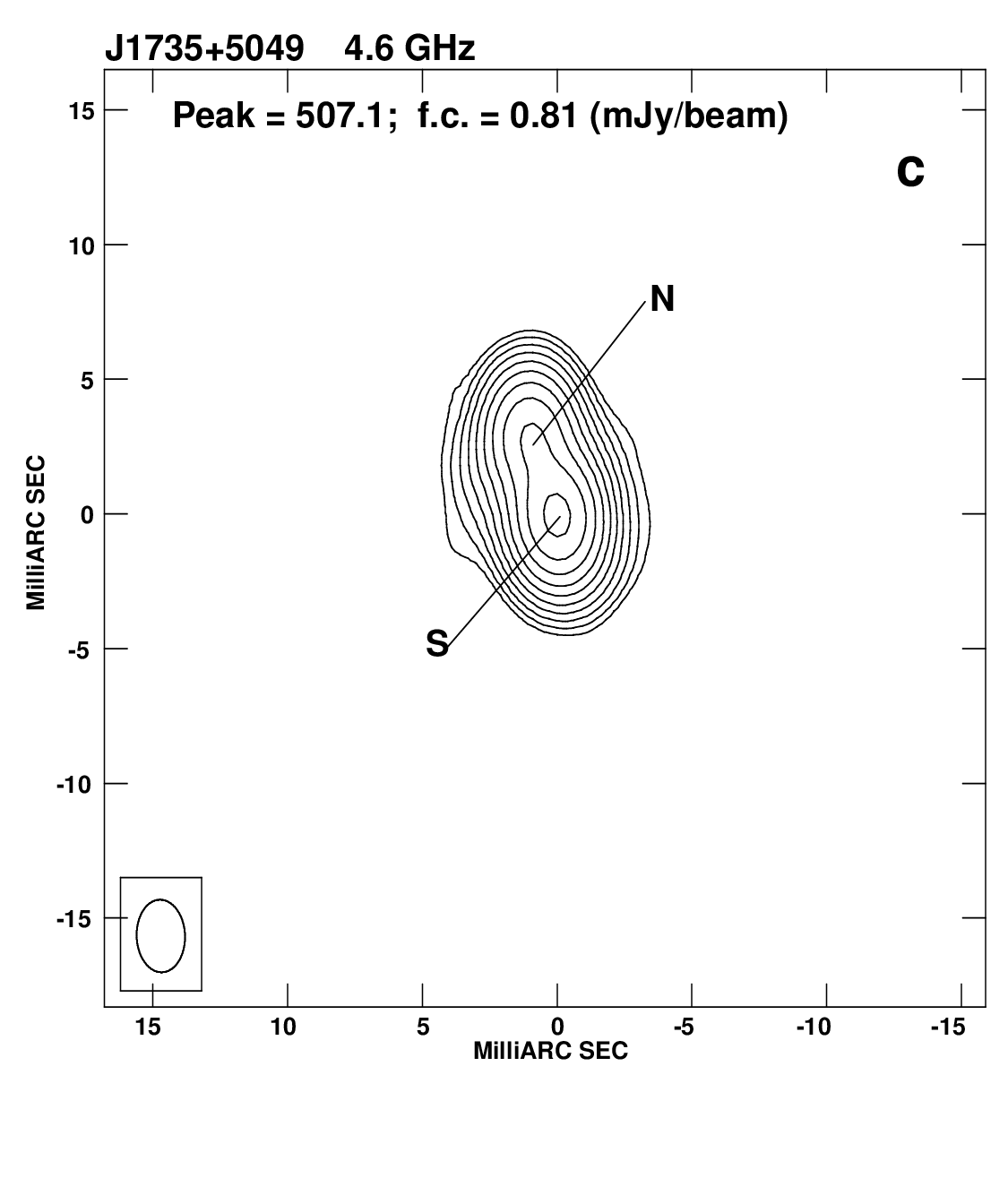}
\includegraphics{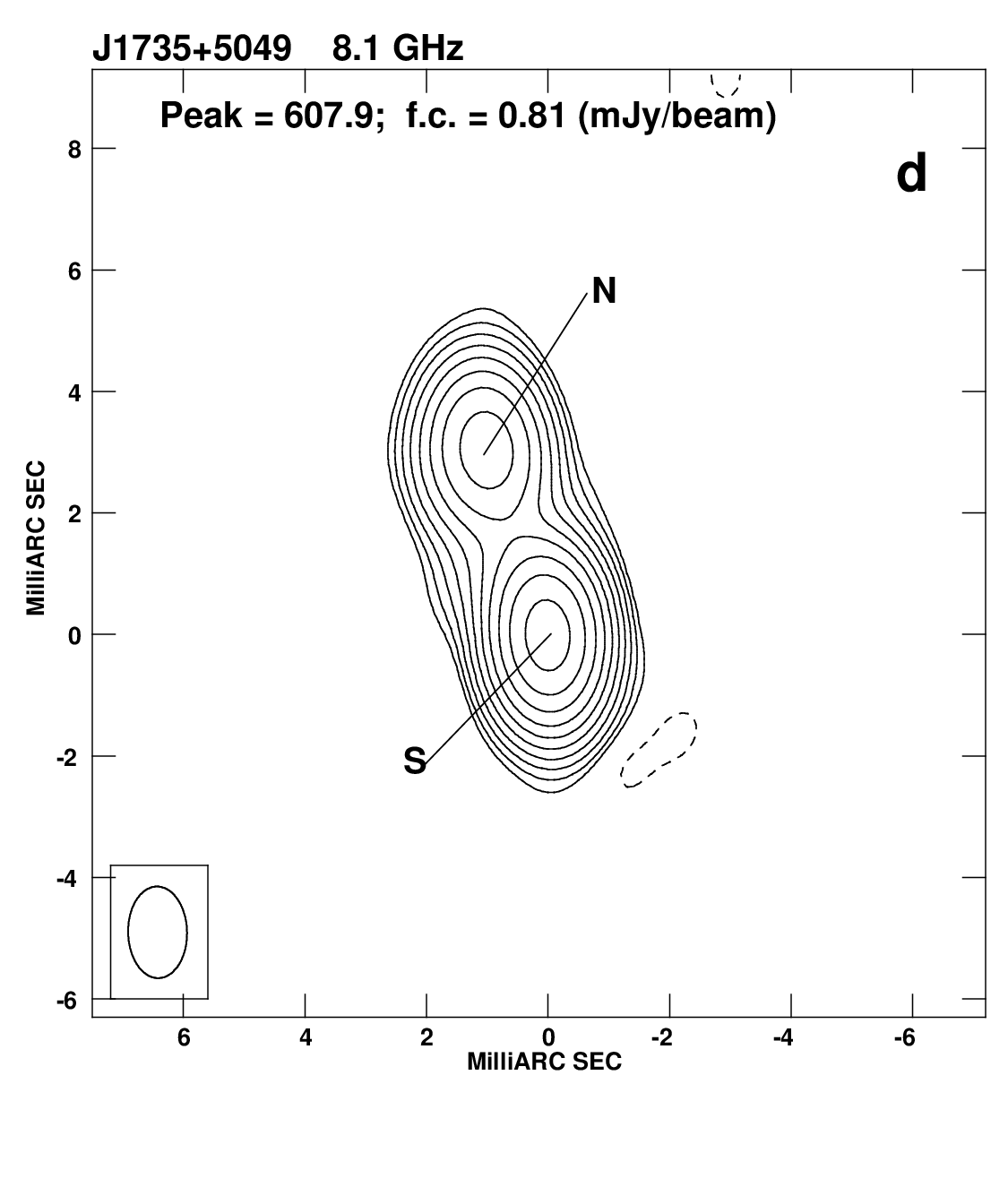}
\includegraphics{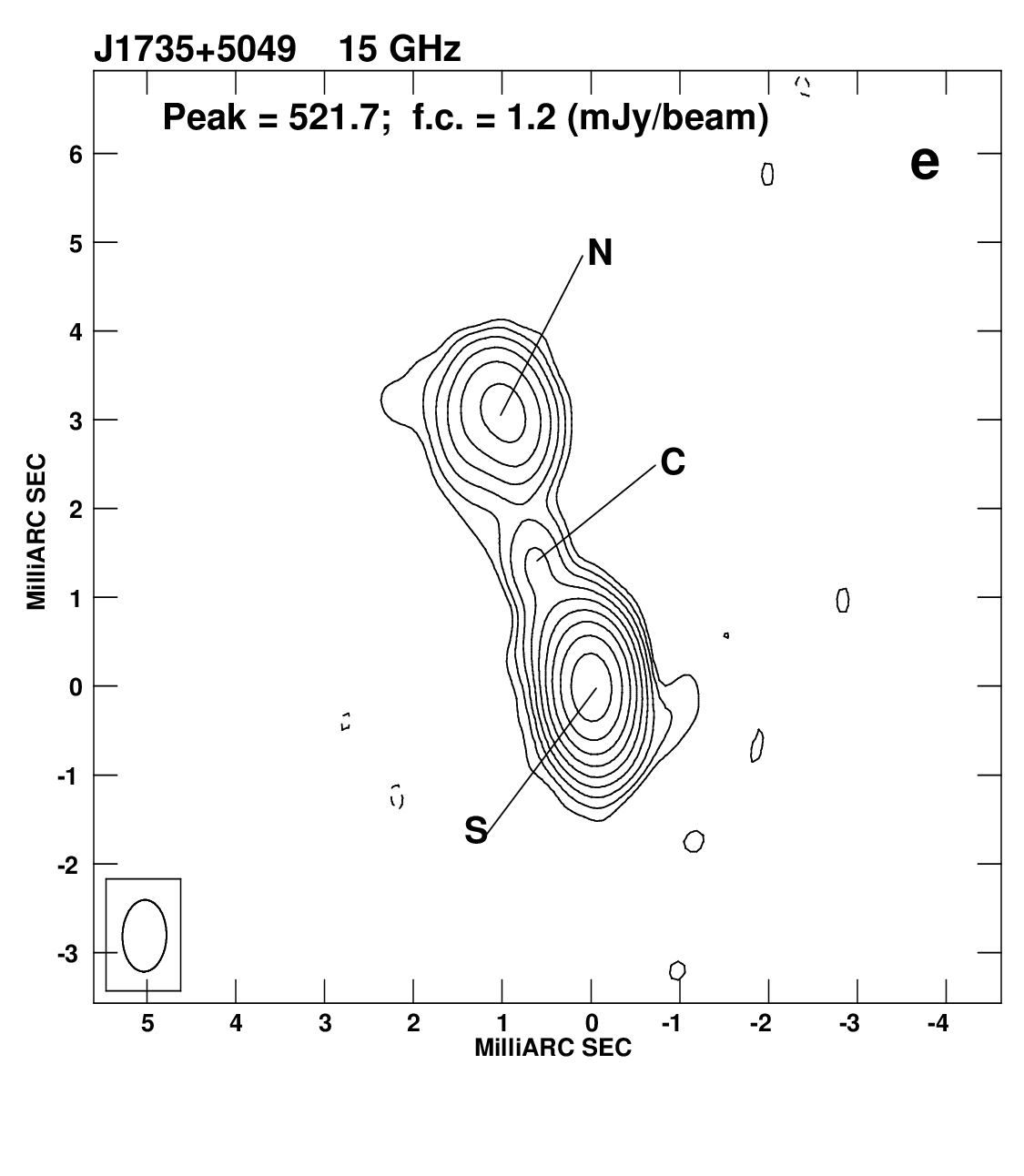}
\includegraphics{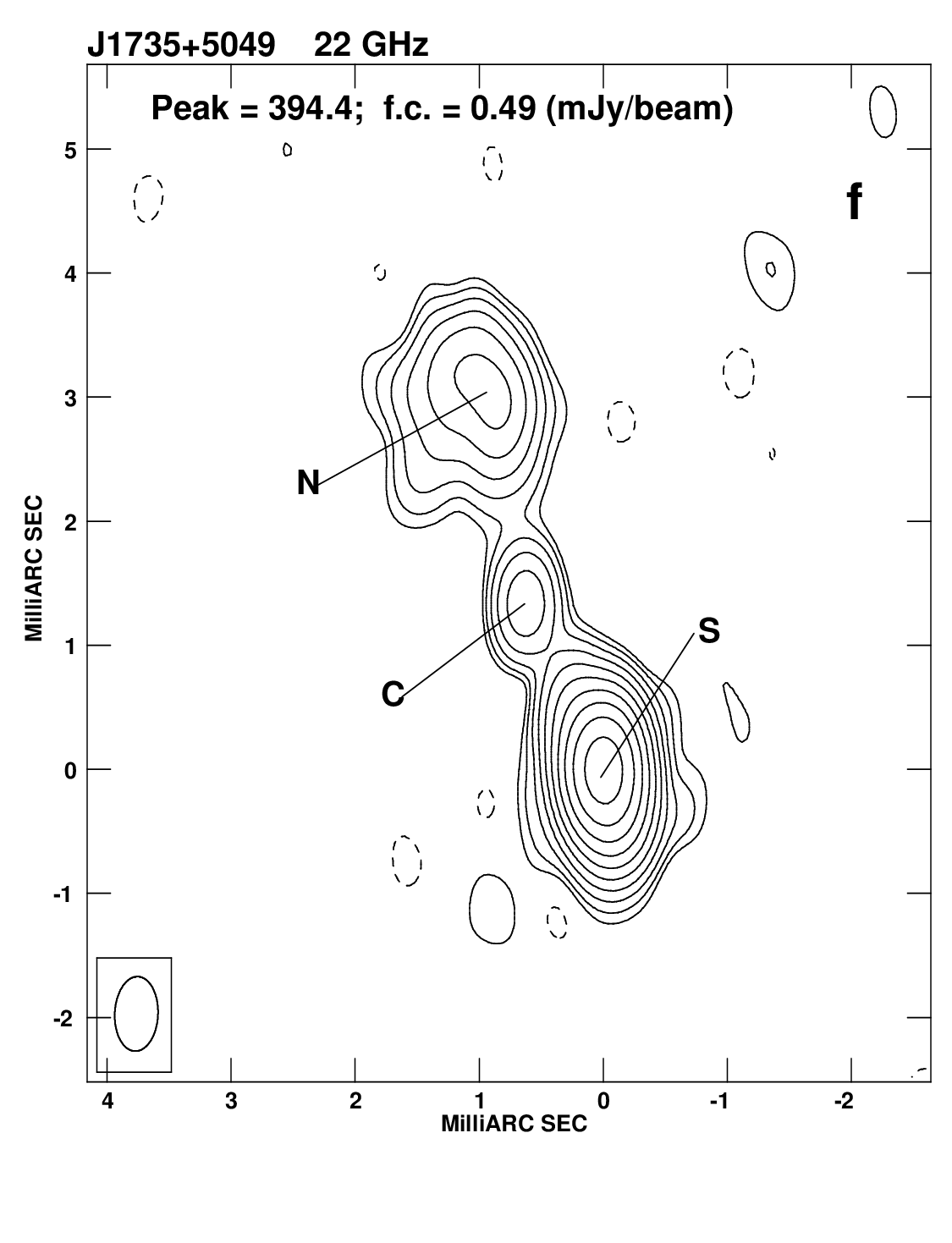}
\vspace{12.5cm}
\caption{VLBA observations of J1735+5049 at 1.6 GHz ({\bf a}), 2.2 GHz
  ({\bf b}), 4.6 GHz ({\bf c}), 8.1 GHz ({\bf d}), 15 GHz ({\bf
  e}), and 22 GHz ({\bf f}). 
On each image we provide the peak flux density in mJy/beam and the first
contour intensity (f.c.) in mJy/beam, which corresponds to three times
the off-source noise level measured on the image plane. Contours
increase by a factor of 2. The restoring beam is plotted in the bottom
left corner of each panel.}
\label{1735}
\end{center}
\end{figure*}

\begin{figure*}
\begin{center}
\includegraphics{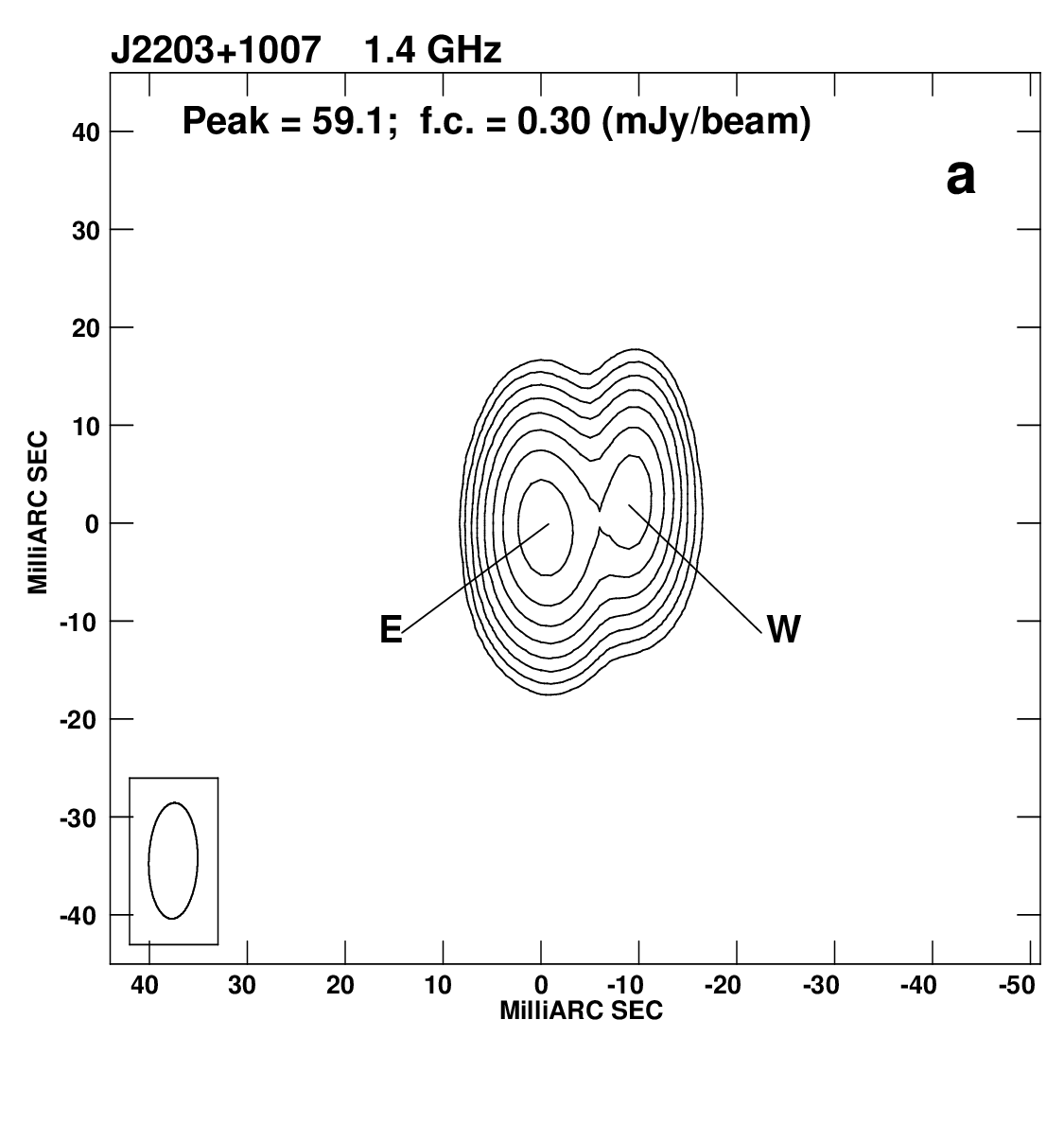}
\includegraphics{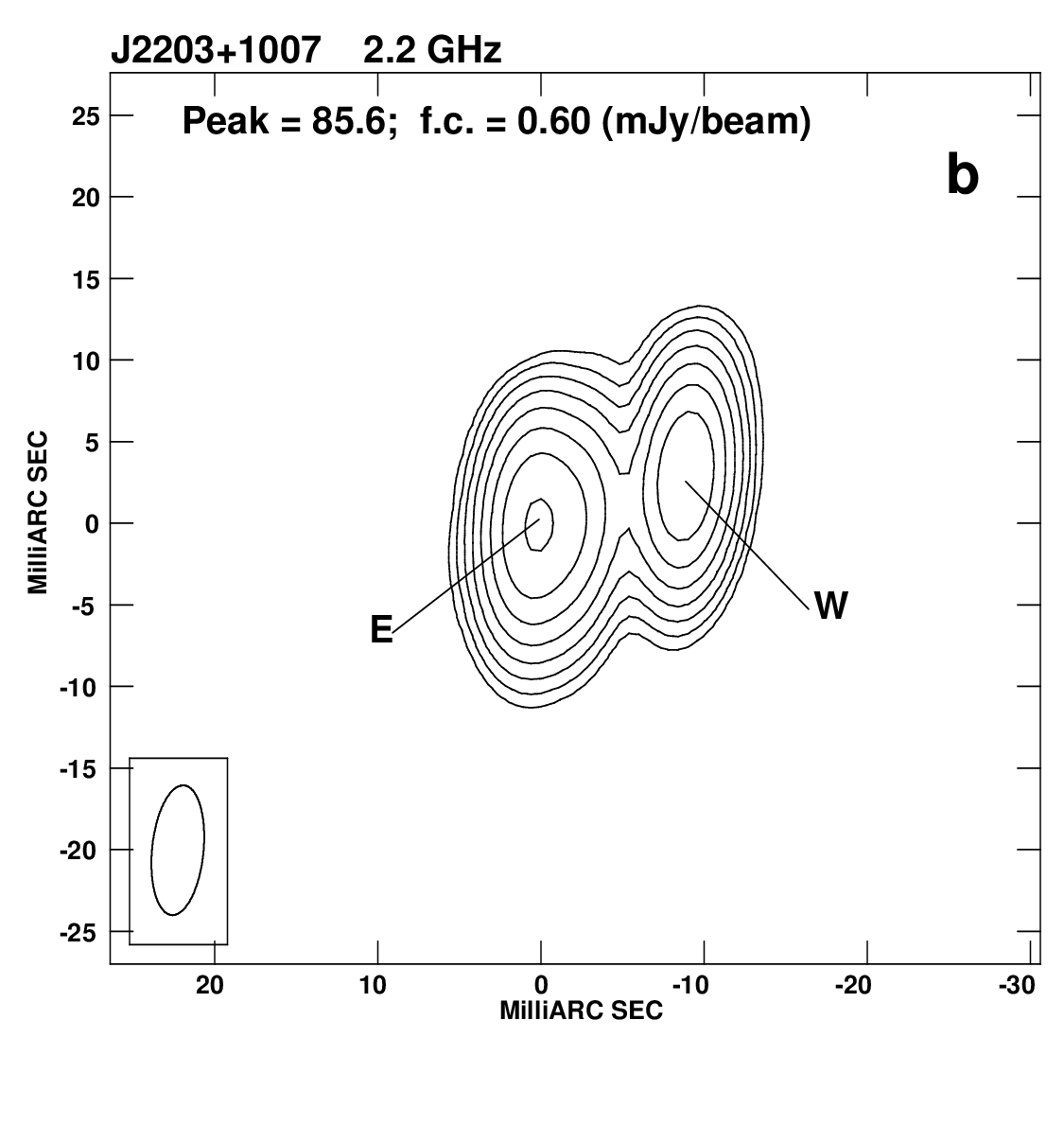}
\includegraphics{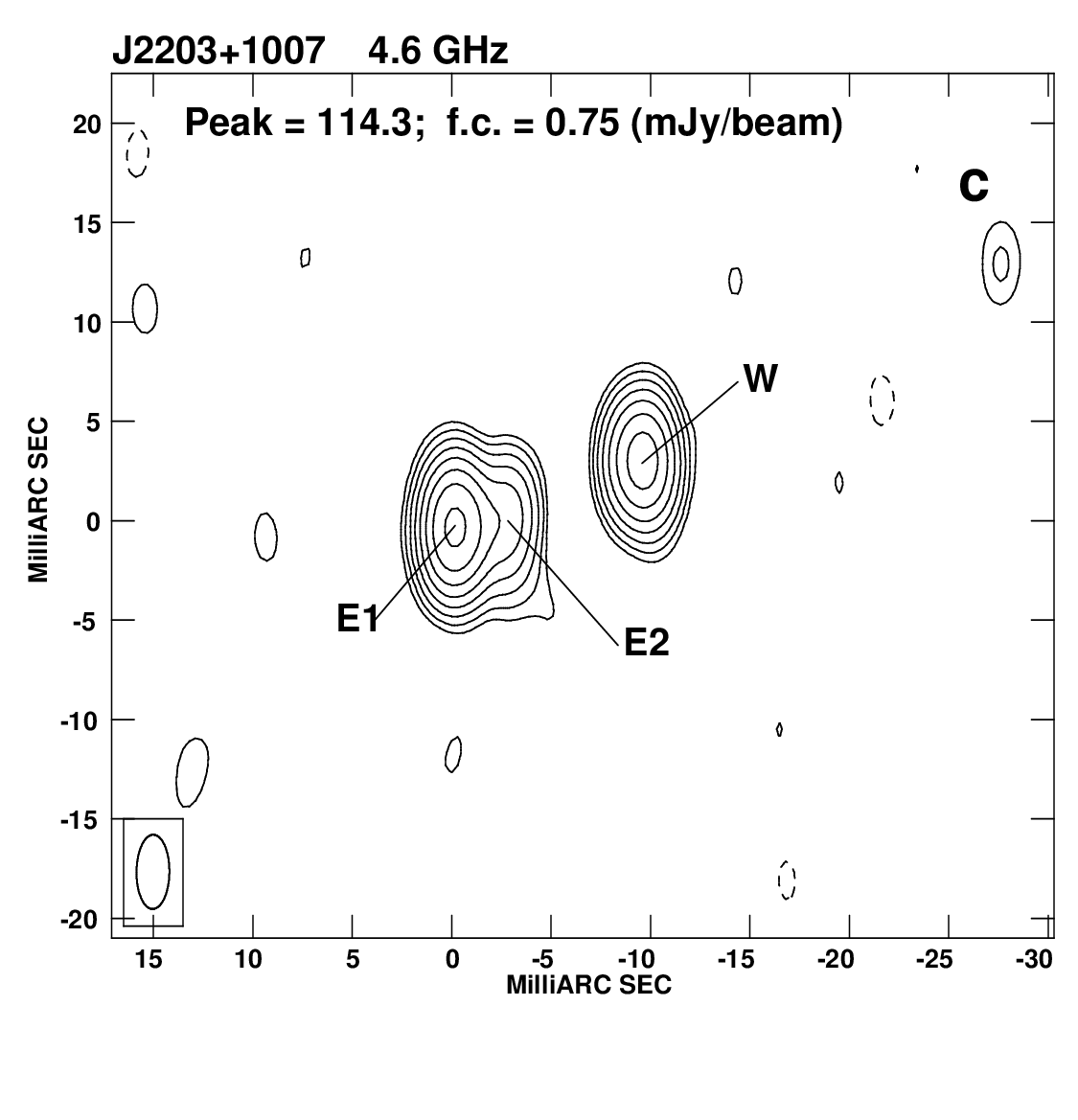}
\includegraphics{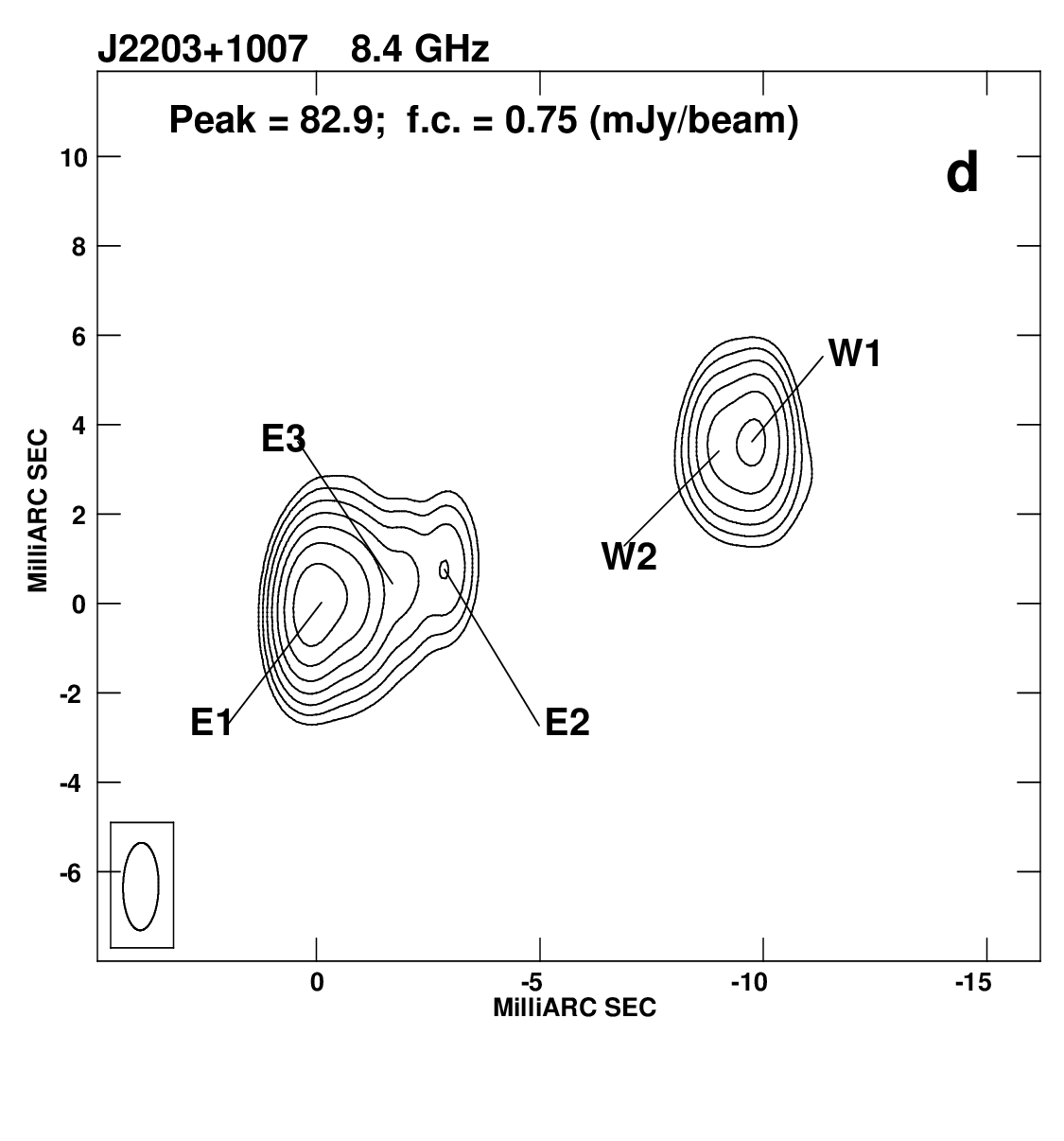}
\includegraphics{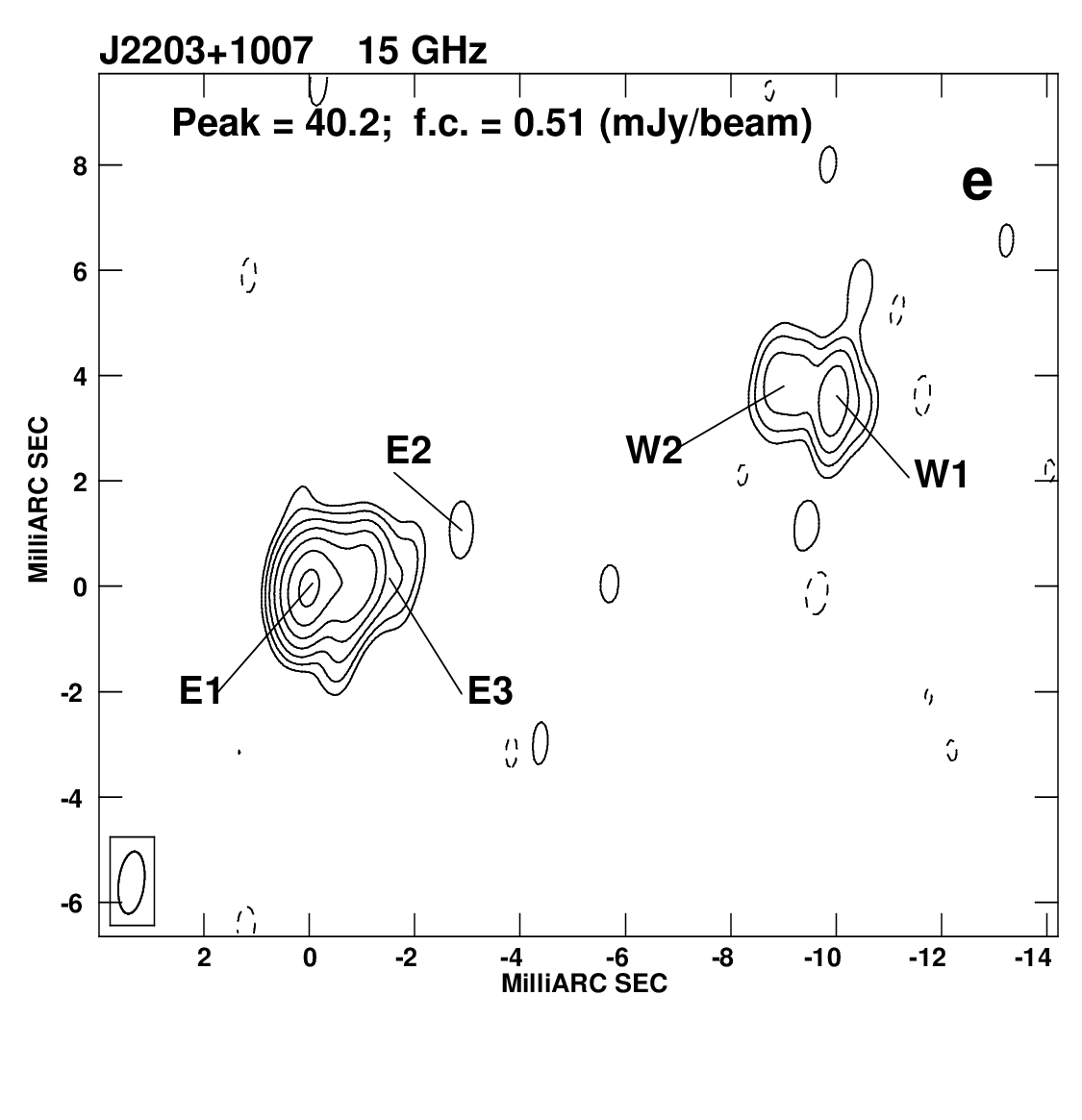}
\vspace{12.5cm}
\caption{VLBA observations of J2203+1007 at 1.4 GHz ({\bf a}), 2.2 GHz
  ({\bf b}), 4.6 GHz ({\bf c}), 8.4 GHz ({\bf d}), and 15 GHz ({\bf
  e}). On each image we provide the peak flux density in mJy/beam and the first
contour intensity (f.c.) in mJy/beam, which corresponds to three times
the off-source noise level measured on the image plane. Contours
increase by a factor of 2. The restoring beam is plotted in the bottom
left corner of each panel.}
\label{2203}
\end{center}
\end{figure*}

\section{Results}

Multifrequency observations with milliarcsecond resolution are the
main tool used to study the morphology and the physical properties 
of very young, compact radio sources.\\
The total flux density of each source is reported in Table
\ref{vlba-tot} and was measured by using TVSTAT which performs an
aperture integration over a selected region on the image plane. We
also derived the flux density and the deconvolved angular size of each
source component by using the \texttt{AIPS} task JMFIT which
performs a Gaussian fit to the source components on the image plane. 
The formal uncertainty on the deconvolved angular size obtained from 
the fit is $<$0.1 mas. As expected, the sum of the flux density 
from each component agrees with the total flux density measured 
on the whole source structure. Source components are referred 
to as north (N), south (S), east (E), west (W), and flat spectrum 
core (C) when detected.  
Observational parameters of the source components are reported in
Table \ref{table-vlba}. \\
The uncertainty on the flux density arises from both the calibration
error $\sigma_{\rm c}$ (see Section 2) and the measurement error
$\sigma_{\rm m}$. The latter represents the off-source rms noise level measured
on the image plane and is related to the source size $\theta_{\rm
  obs}$ normalized by the beam size $\theta_{\rm beam}$ as
$\sigma_{\rm m}$=
rms $\times$ ($\theta_{\rm obs}$/$\theta_{\rm beam}$)$^{1/2}$. 
The flux density error $\sigma_{\rm S}$ reported in Tables
\ref{vlba-tot} and \ref{table-vlba} 
takes into account both uncertainties.
It corresponds to $\sigma_{\rm S} = \sqrt{ \sigma_{\rm c}^{2} +
  \sigma_{\rm m}^{2}}$, and is generally dominated by $\sigma_{\rm
  c}$ which usually exceeds $\sigma_{\rm m}$.\\

\subsection{Notes on individual sources}

{\bf J0037+0808}: no optical counterpart is found in the Sloan Digital
Sky Survey Data Release 9 (SDSS DR9) image
\citep{ahn12}.  
Observations at the higher frequencies resolve the source
  structure into three main components (Fig. \ref{0037}).
The East component, E, dominates the radio emission,
accounting for more than 80 per cent of the total flux density. 
At high frequencies component E is resolved into two sub-components, E1 and
E2, separated by about 1 mas. Component E2 has a flat
spectrum $\alpha_{\rm 8.4}^{\rm 15} \sim 0.25$ and it may host the
source core. If E2 is the source core, the source structure is
symmetric, with component E1 and W roughly at the same distance from
E2. However, their flux densities are very different, and the flux
density ratio is $S_{\rm E1}/S_{\rm W} \sim$ 4.4 and 10 at 8.4 and 15
GHz, respectively.  \\ 
The source has a total angular size of about 3.4 mas.\\

{\bf J1335+5844}: in the SDSS image this source is associated with a galaxy
with an
estimated photometric redshift of 0.58 \citep{ahn12}. At low frequencies the source
shows a double structure with the South component, S, being the
brightest one (Fig. \ref{1335}). 
On the contrary, above 8.1 GHz the radio emission is
dominated by the northern component, N, which is resolved into three
sub-components: N1 is a very compact hotspot, while N2 and N3 are likely jet
regions. The diffuse emission enshrouding the southern hotspot is
almost resolved out in the images at high frequencies.\\
Component C has a flat spectrum ($\alpha_{\rm 15}^{\rm 22} \sim$0)
and is identified as the source core. Its flux density is 1.6 mJy at
22 GHz, representing about 0.5 per cent of the total flux density at
such frequency.  
The core is located at about 5.6 mas ($\sim$37 pc)
from N1 and 7.6 mas ($\sim$50 pc) from S, indicating that the brighter
component at the higher frequencies is also the closest to the
core. \\
The core component was not detected in earlier VLBA
observations presented in \citet{mo06} likely due to the inadequate
dynamic range, while it
was detected in deeper 15-GHz VLBA observations by
\citet{dd05} and \citet{an12}, 
but the lack of simultaneous spectral index information precluded a
secure identification as the nucleus. \\
The source has a total angular size of about 15.5 mas, which
corresponds to a linear size of $\sim$100 pc.\\
J1335+5844 was not part of the COINS sample of CSO candidates selected
by \citet{peck00}
due to its flat overall spectrum between 1.4 and 5 GHz
\citep{taylor95}. However, the flat spectrum does not come from a
boosted core-jet source, but it arises from the
northern, compact hotspot which is absorbed below 10 GHz (Fig. \ref{1335}).
The unambiguous detection of the core allows us to classify this
source as a genuine CSO.\\ 

{\bf J1735+5049}: the source is associated with a galaxy at
redshift z=0.835 \citep{britzen08}. At low frequencies
the radio source shows a double structure, and the radio emission is
dominated by the southern component S. In the 15-GHz image 
a central component is detected almost halfway 
between component N and S, becoming
clearly resolved at 22 GHz (Fig. \ref{1735}). 
This component has an inverted spectrum ($\alpha_{\rm 15}^{\rm 22}
\sim$-0.4) and is
identified as the source core. Its flux density is 14 mJy at 22 GHz,
representing 2 per cent of the total flux density at such frequency.\\
The source has a total angular size of about 5.3 mas, which
corresponds to a linear size of $\sim$40 pc.\\
J1735+5049 was not part of the COINS sample of CSO candidates 
selected by \citet{peck00} due to its flat overall spectrum between 1.4 and 5 GHz 
\citep{taylor95}. As in the case of J1335+5844, 
the flat spectrum does not come from a
boosted core-jet source, but it arises from the
compact hotspots which are absorbed below 6 GHz (Fig. \ref{1735}). The
unambiguous detection of the core allows us to classify this
source as a genuine CSO.\\

{\bf J2203+1007}: the source is associated with a narrow line galaxy at
redshift z=1.005 \citep{healey08}. The radio emission has a double
structure (Fig. \ref{2203}), and 
is dominated by the East component, E, which accounts for
more than 65 per cent of the total flux density. Observations at the higher
frequencies resolve components E and W into sub-components,
showing the complex morphology of the object. No flat-spectrum
component is detected, leaving the core region unidentified at the mJy
level, limiting its fractional contribution to the total flux density
below 1 per cent at 15 GHz.\\
The source has a total angular size of about 12 mas, which
corresponds to a linear size of $\sim$97 pc.\\

\subsection{Radio spectra}

\begin{figure*}
\begin{center}
\includegraphics{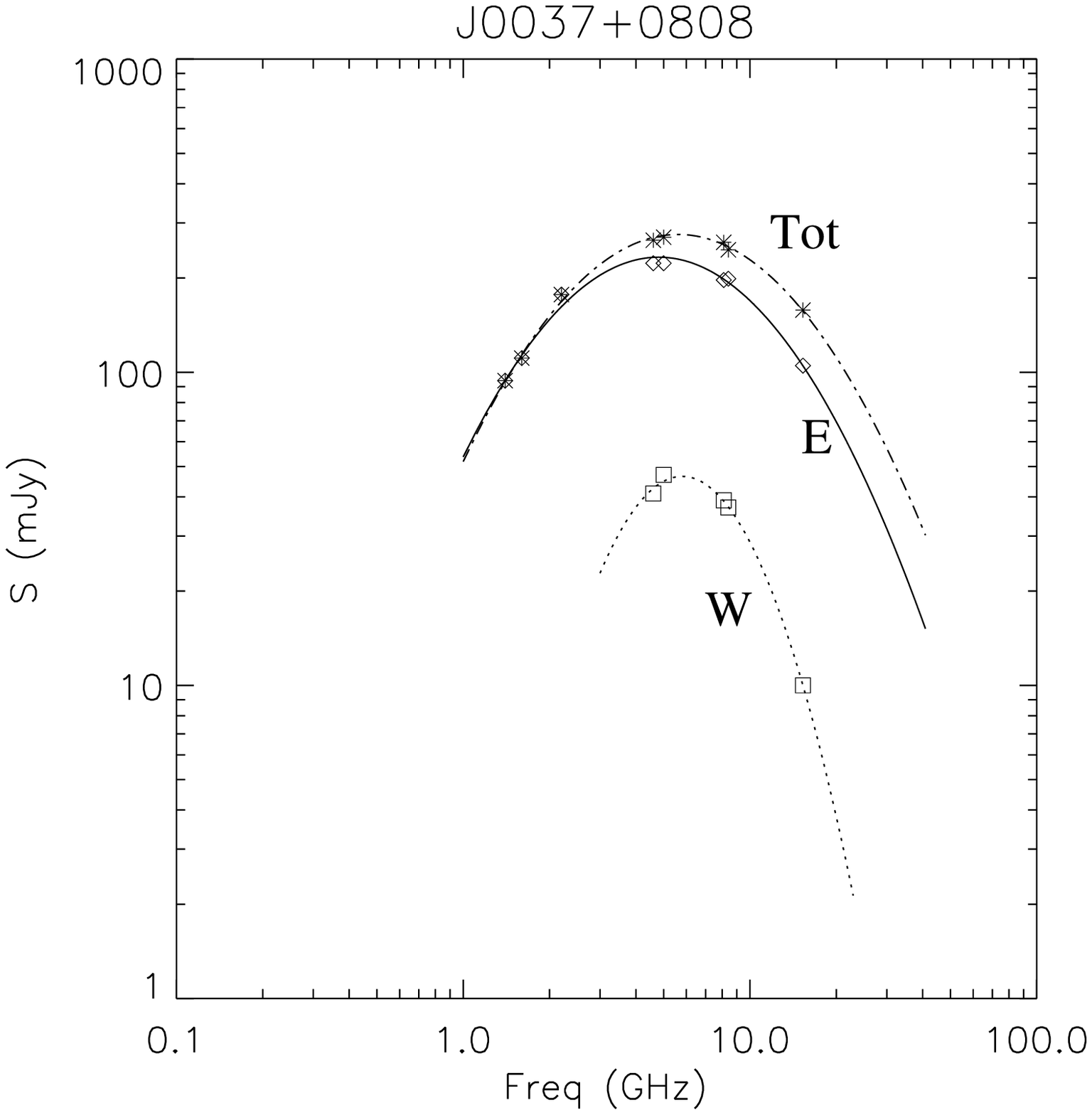}
\includegraphics{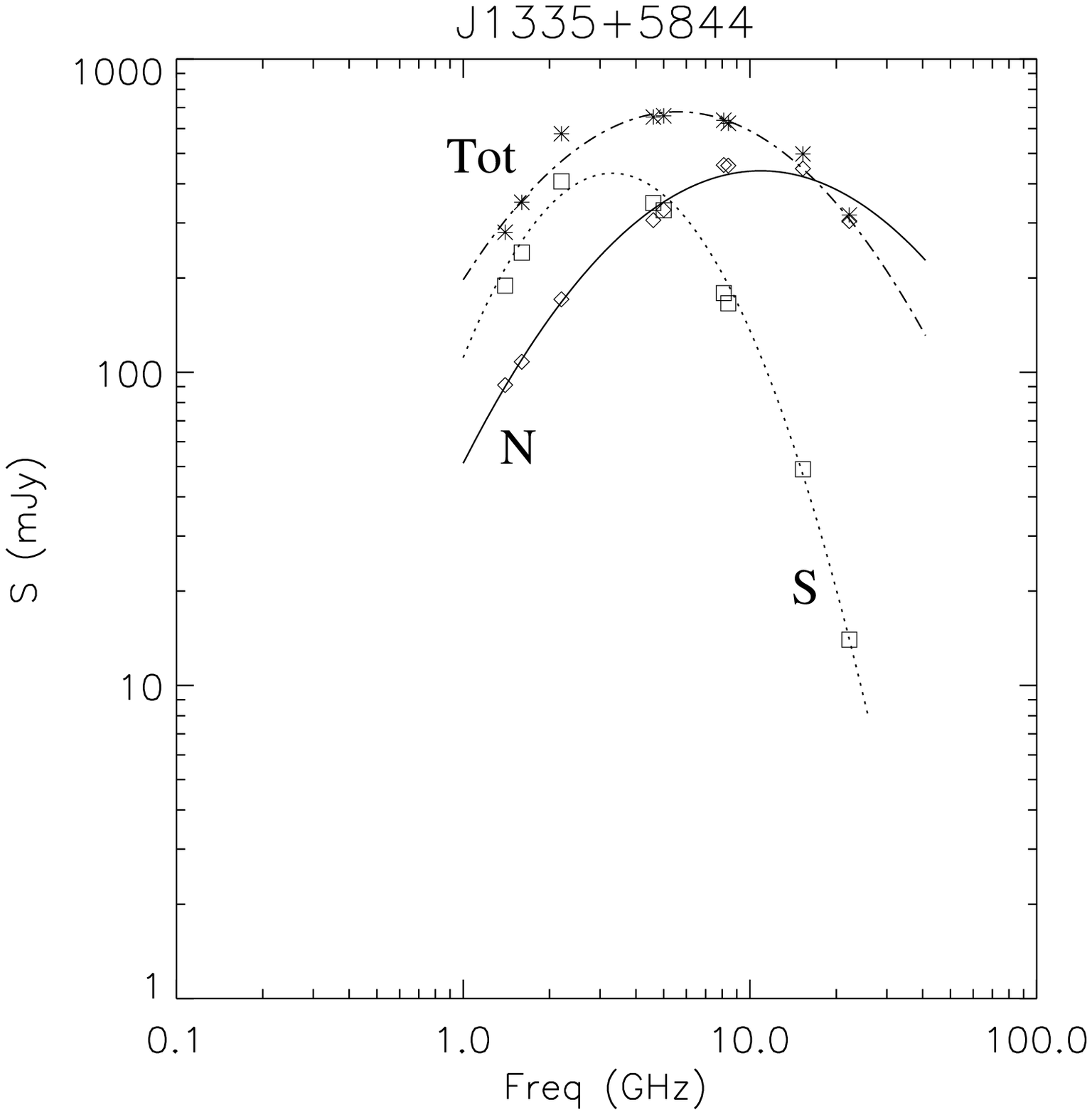}
\includegraphics{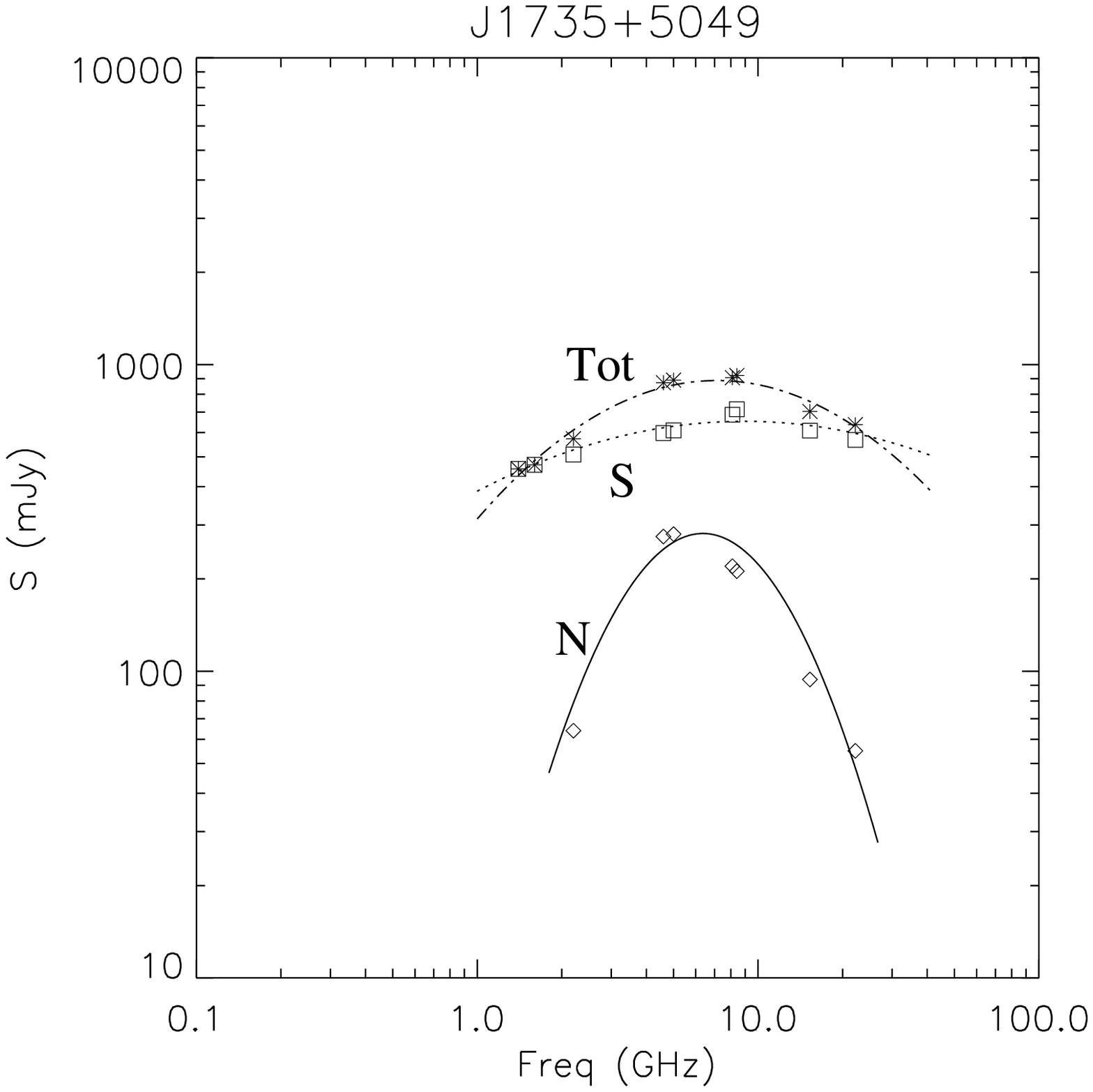}
\includegraphics{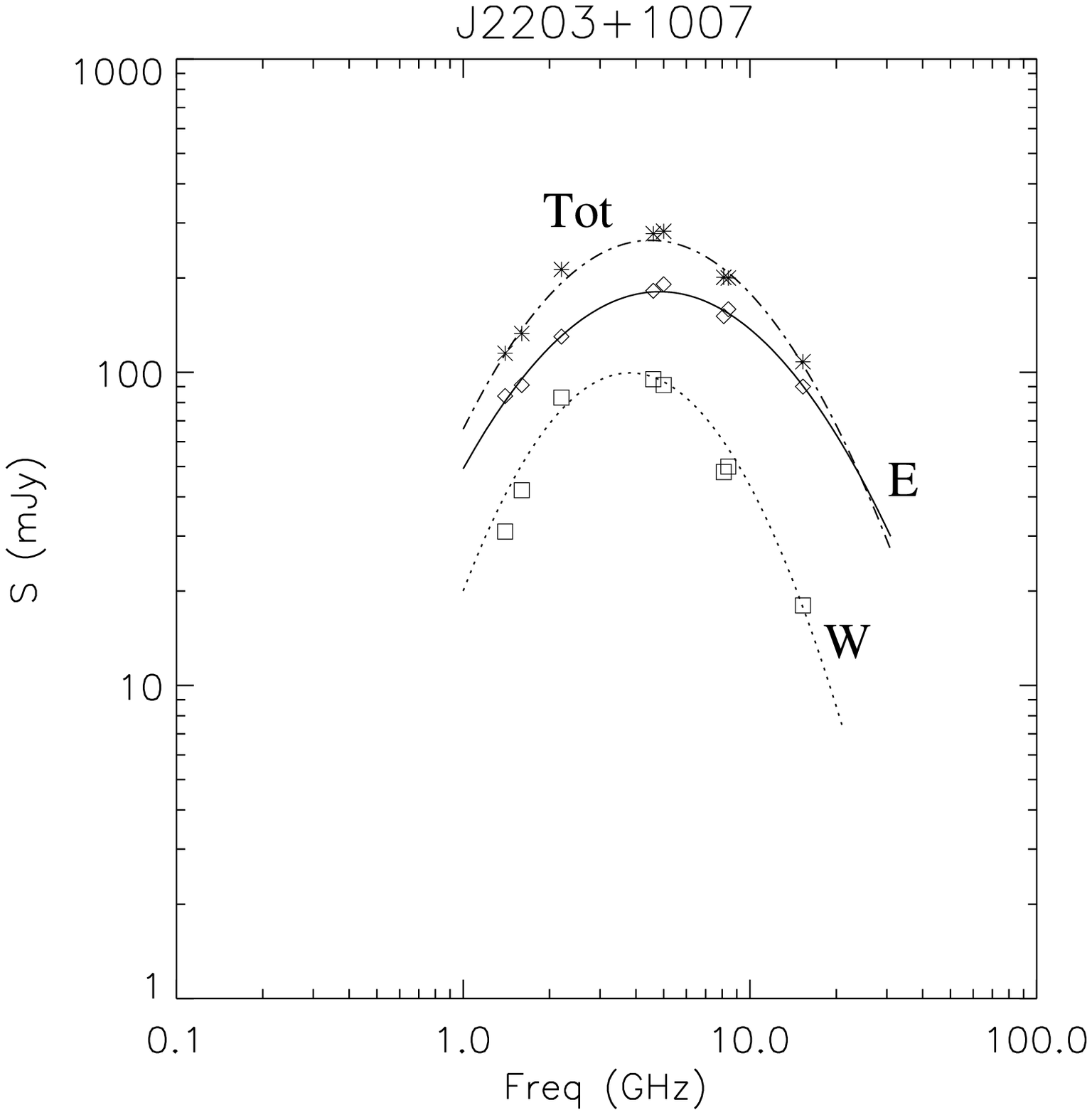}
\vspace{6cm} 
\caption{Spectra of the HFP sources and of their main components. }
\label{spectra}
\end{center}
\end{figure*}

The availability of high resolution observations at various
frequencies performed almost simultaneously in both the
optically-thick and -thin regimes allows us to study the radio
spectrum of the main components. Following \citet{mo07} and
\citet{mo10} we modelled each spectrum with a pure analytical
function:

\begin{displaymath}
Log S = a + Log (\nu) \times (b + c \times Log(\nu))
\end{displaymath}

\noindent where $S$ is the flux density, $\nu$ is the frequency, and
$a$, $b$, and $c$ are numeric parameters without any direct physical
meaning. The best fits to the spectra are shown in Fig. \ref{spectra}, 
and the derived peak frequency and peak flux density
are reported in Table \ref{tab_peak}. Tabulated uncertainties 
are from the fit only.

\begin{table}
\caption{Spectral parameters of the source components. 
Column 1: source name; Column 2: source component; Column 3:
  peak frequency; Column 4: peak flux
  density; Columns 5 and 6:
  spectral index computed in the optically thick and optically thin
  part of the spectrum, respectively.}
\begin{center}
\begin{tabular}{cccccc}
\hline
Source&Comp&$\nu_{\rm p}$&$S_{\rm p}$&$\alpha_{b}$&$\alpha_{a}$\\
      &    &  GHz       &  mJy     &  &  \\
(1)   & (2)& (3) & (4) & (5) & (6) \\
\hline
&&&&&\\
J0037+0808&E& 4.8$\pm$0.5& 233$\pm$2& -0.7 &0.7  \\
          &W& 5.8$\pm$1.0&  47$\pm$4&  - & 1.4 \\
          &Tot&5.6$\pm$0.5& 276$\pm$2& -0.8 & 0.4\\ 
J1335+5844&N&10.9$\pm$2.0&440$\pm$7& -0.9 & 0.4\\
          &S& 3.3$\pm$0.5&432$\pm$10& -1.7 & 1.4\\
          &Tot&5.6$\pm$0.5& 679$\pm$2& -0.7 & 0.5\\
J1735+5049&N& 6.3$\pm$0.5& 271$\pm$2& -2.0 & 1.1\\
          &S& 8.9$\pm$0.5& 653$\pm$1&  -0.2& 0.2\\
          &Tot& 7.1$\pm$1.0& 887$\pm$6& -0.4 & 0.4\\
J2203+1007&E & 4.8$\pm$0.6 & 181$\pm$2& -0.6& 0.7\\ 
          &W & 3.8$\pm$0.5& 100$\pm$3&  -0.9& 1.4\\
          &Tot& 4.5$\pm$0.5& 264$\pm$3& -0.7 & 0.8\\
&&&&&\\
\hline
\end{tabular}
\end{center}
\label{tab_peak}
\end{table}

\begin{table}
\caption{Physical parameters of the source components. 
Column 1: source name; Column 2: source component; Column 3:
  radio luminosity; Column 4: volume; Columns 5 and 6: equipartition
  magnetic field and the magnetic
  field from observational parameters, respectively (see Section 4.2).}
\begin{center}
\begin{tabular}{cccccc}
\hline
Source&Comp&L& V& $H_{\rm eq}$& $H$\\
      &    &  erg s$^{-1}$& cm$^{3}$ &  mG&  mG \\
(1)   & (2)& (3) & (4) & (5) & (6) \\
\hline
&&&&&\\
J0037+0808&E&9.3$\times$10$^{42}$& 1.4$\times$10$^{54}$& 86 & 9  \\
          &W& 2.2$\times$10$^{42}$&  1.4$\times$10$^{54}$&  57& 55 \\
J1335+5844&N&3.9$\times$10$^{44}$&2.4$\times$10$^{55}$& 111 & 268 \\
          &S& 1.5$\times$10$^{44}$& 4.5$\times$10$^{57}$& 19 & 55\\
J1735+5049&N& 3.2$\times$10$^{44}$& 7.9$\times$10$^{56}$& 39 & 1800 \\
          &S& 1.7$\times$10$^{45}$& 2.6$\times$10$^{55}$&  165 & 37\\
J2203+1007&E & 3.7$\times$10$^{44}$ & 3.2$\times$10$^{56}$& 52 & 200\\ 
          &W & 1.4$\times$10$^{44}$& 8.2$\times$10$^{56}$&  30& 988\\
&&&&&\\
\hline
\end{tabular}
\end{center}
\label{magnetic}
\end{table}

\noindent The peak parameters derived from the integrated spectra are in good
agreement with the values obtained in \citet{mo07} on the basis of VLA
observations, pointing out both the 
lack of significant spectral variability and the lack of extended low-surface
brightness emission undetectable by VLBA observations. The overall spectra peak
at about 5 GHz or at higher frequencies, as a consequence of the selection
criteria \citep[see][]{dd00}. Fig. \ref{spectra} shows the
overall spectrum together with the spectra of the main components as
derived by our VLBA observations. In all the four sources, 
the overall spectra are easily explained as the result of the
superposition of the spectra of the individual main components, in
general two, both characterized by different peak frequencies. 
When the main radio
emitting regions have different turnover frequencies, their relative
contribution to the overall spectrum may change significantly at the
different frequencies.
As a consequence, the peak frequency
of the overall spectrum is located between the spectral peaks of the
dominant sub-components.
This is particularly remarkable in the case of J1335+5844
where the two main components have similar peak flux densities, but very
different peak frequencies: 3.3 and 10.9 GHz. As a result, the peak of the overall
spectrum occurs at about 5.6 GHz. \\   
In the other sources the overall spectrum is mainly influenced by the
component which generally dominates the radio emission at all the observing
frequencies. \\

\section{Discussion}

\subsection{Radio morphology}

Our multifrequency, high angular resolution VLBA observations
allow us to resolve the source structure in many compact components
and to determine their nature by using the spectral index
information. In general the radio emission is dominated by one of the 
hotspots/lobes. The flux density ratio between the components of the same
radio source ranges between 2 and 10, showing an asymmetric
luminosity distribution. 
The optically-thin 
spectral index of the hotspot components is about 0.7, although it is
difficult to disentangle the hotspot region from the surrounding
mini-lobe. 
In J1335+5844 and
J1735+5049, the main
hotspot is very compact and hence
its radio emission is self-absorbed below 10 GHz. For this reason
the spectra are flat ($\alpha = 0.2 -
0.4$) due to the presence of opacity effects up to high frequencies
which prevent us from determining the intrinsic spectral index of the
fully optically-thin emission.\\
The steep
spectra found in other compact components ($\alpha > 1.0$) may reflect
some missing flux density at the highest frequency likely due to the
lack of the shortest baselines.\\ 
Among the four HFPs studied in this paper, the core is unambiguously
detected in J1335+5844 and J1735+5049, and represents a small
fraction ($\leq$2 per cent) of the source total emission at the highest
observing frequency indicating the lack
of boosting effects. In J0037+0808
we suggest the central component, E2, as the core candidate on the basis of
its flat spectrum. In this source E2 component accounts for about 27
per cent of the total flux density at the highest frequency. However,
the lack of significant variability and the small source size may
suggest that E2 is likely a blend of the true source core and the
jet base. \\
No core candidate is found in J2203+1007. \\
In J0037+0808 and J1735+5049, the core component is located roughly
midway between the outer components. On the other hand, in J1335+5844
the compact and most luminous hotspot is the one closest to the core
component, which is opposite to what is expected from projection and
relativistic time dilation effects. 
Such asymmetry is found in a relatively large
fraction ($\sim52$ per cent)  
of GPS/CSS sources \citep{dd13}, and suggests a strong interplay
between the radio emission and the environment. 
The very compact size of the hotspot in J1335+5844 and J1735+5049 
is likely due to the interaction 
between the radio emission and an inhomogeneous, dense medium, which
confines the component by reducing the adiabatic expansion, and
enhances its luminosity by amplifying the radiative losses. The presence of an
inhomogeneous ambient medium enshrouding young radio sources was found
in the HFP radio galaxies J0428+3259 and J1511+0518, where free-free
absorption from an ionized medium was detected only in front of one of the
two lobes \citep{mo08}.  \\

\subsection{The magnetic field}

The direct measurement of the magnetic field from observable
  quantities is a difficult task to perform.
A way to bypass the problem is assuming that a radio
source is in minimum energy conditions, which means that the energy
stored in relativistic particles is approximately equal to the
magnetic energy. The equipartition magnetic field $H_{\rm eq}$ is then
obtained by:

\begin{equation} 
H_{\rm eq} = \left( \frac{c_{12} L}{V} \right)^{\frac{2}{7}}
\label{equi}
\end{equation}

\noindent where $L$ is the radio luminosity, $V$ is the volume and
          $c_{12}$ is a constant which is tabulated in \citet{pacho70}
and depends on the spectral index and
          the upper and lower cutoff frequencies. In the case of the HFP
          components, we assume an average spectral index $\alpha
          =0.7$, a lower cutoff frequency, $\nu_{1}$ = 10 MHz, and an
          upper cutoff frequency, $\nu_{2}$ = 100 GHz. The radio
          luminosity is obtained by:

\begin{equation}
L = \frac{4 \pi D_{\rm L}^2}{(1+z)^{1- \alpha}}  \int_{\nu_{1}}^{\nu_{2}} S(\nu) d
  \nu
\label{luminosity}
\end{equation}

\noindent We approximate the volume of the source components
to a prolate ellipsoid which is homogeneously filled by
the relativistic plasma: 

\begin{equation}
V = \frac{\pi}{6} \left( \frac{D_{\rm L}}{(1+z)^2} \right)^3
\theta_{\rm maj} \theta_{\rm min}^2 \,\, .
\label{volume}
\end{equation}

\noindent where $D_{\rm L}$ is the luminosity distance,
  $\theta_{\rm maj}$ and $\theta_{\rm min}$ are the major 
and minor angular size, respectively, and $z$ is 
the redshift.  
If in Eq. \ref{equi} we consider the luminosity and volume derived
from Eqs. \ref{luminosity} and \ref{volume}, we obtain
equipartition magnetic fields ranging between 20 mG to 180 mG (see
Table \ref{magnetic}), which are
similar to the values derived for the components of other HFP sources
studied earlier \citep{mo08, mo12}. \\
If the spectral peak is produced by synchrotron
self-absorption (SSA), we have an independent way to compute the magnetic
field by using observable quantities only.
In this case the magnetic
field, $H$, in Gauss, is:

\begin{equation}
H = \frac{\nu_{p}^5 \theta_{\rm maj}^2 \theta_{\rm min}^2}{f(
  \alpha)^5 S_{p}^2 (1+z)}
\label{ssa}
\end{equation}

\noindent where $\nu_{p}$ is the peak frequency in GHz, $S_{p}$ is the peak
flux density in Jy, $\theta_{\rm maj}$ and $\theta_{\rm min}$ are the major
and minor angular size in milliarcsecond 
of the source component respectively,
$f(\alpha)$ is a function that depends slightly on $\alpha$, and it
is $\sim$8 \citep[see e.g.][]{kellermann81,odea98}. \\
The magnetic fields are computed 
for the source components with a well-sampled spectrum and hence
  with an accurate estimate of the peak parameters.
In Equation \ref{ssa} we consider the peak flux density and peak
frequency obtained by the fit to the spectra (Table \ref{tab_peak}), and the
angular sizes measured from the images (Table
\ref{table-vlba}). Following the approach by \citet{read94} 
we consider component angular sizes that are 1.8 times larger than the
full width at half maximum derived by the Gaussian fit. In the case of
the southern component of J1335+5844, the component size was measured
directly from the image due to the complex structure of the emitting
region. \\
The magnetic fields obtained for the source components range between a
few mG up to 1.8 G (Table \ref{magnetic}). In half of the components,
such values 
usually agree with the equipartition 
magnetic fields within a factor of $\sim$2-4. In the other
components the difference between the magnetic field values obtained
with the two approaches may be larger than an
order of magnitude. In the case of the northern component of
J1735+5049, and both the components of J2203+1007, the magnetic field
computed from the peak parameters is much higher than the
equipartition magnetic field. In the work by \citet{mo08}, such a large
difference was found in one of the components of J0428+3259 and
J1511+0518, where the spectral peak was interpreted as due to free-free
absorption (FFA) by ionized thermal medium, rather than in terms 
of pure SSA. This explanation
was supported by the analysis of the optically-thick part of the spectra
which could be accurately described by a FFA model only.\\
In the case of J1335+5844 and J1735+5049, the large values of the magnetic field
seems to reflect a non-homogeneous synchrotron component. The angular
size estimated from the images refers to blended components
that cannot be adequately resolved by these VLBA
observations. Therefore, the
estimated angular size should be considered an upper limit.
As a consequence, the magnetic field derived
from observational parameters is largely overestimated because it
strongly depends on the angular
size ($H \propto \theta^4$, while $H_{\rm eq} \propto \theta^{-6/7}$). The
optically-thick spectral index $\alpha_{b} > -2.5$ supports the
inhomogeneity of the source components, rather than the presence of FFA.\\

\subsection{Physical properties and the radio source evolution}

\begin{figure}
\begin{center}
\includegraphics{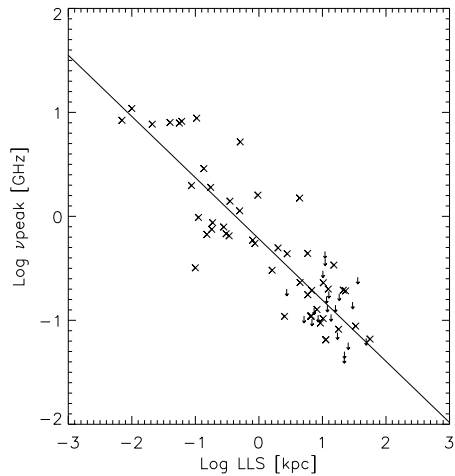}
\vspace{6.5cm}
\caption{Rest-frame peak frequency versus largest linear size for the
  sources of the selected sample (Table \ref{samplone}). 
The solid line represents the best
  linear fit (see Section 4.3). Arrows represent upper limit to the
  peak frequency.}
\label{peak}
\end{center}
\end{figure}

In the context of the evolutionary models, it is important to
constrain the physical properties of young radio sources and
how they evolve in time/size. However, observational information must arise from
samples of genuinely young radio sources, without the contamination of
other populations of objects, like blazars which may temporarily match the
selection criteria of GPS/HFP samples \citep[see
  e.g.][]{tinti05,torniainen05,mo07,hancock10}. \\
For a correct identification of the nature of the radio source, it is
important to have the proper classification of the radio
morphology. However, this is most difficult in the smallest objects
where self-absorption changes the source morphology with
frequency (see e.g. J1335+5844, Section 3.1). 
The secure detection of the core component is a crucial requirement for
classifying a source as a genuine CSO. \\
With the aim of determining the
physical properties during the early stage of the radio emission,
we built a sample of {\it bona-fide} young
radio sources with an unambiguous detection of the core region 
(Table \ref{samplone}). 
The sources were selected from HFP, GPS and CSS samples by 
\citet{dd00,cstan09,dd95,cstan98,peck00,cfanti01,rfanti90,snellen98} in order
to span a wide range of linear size, from a few pc up to tens kpc,
for high luminosity radio sources. 
To derive the intrinsic physical parameters, we selected only the
sources with information on the redshift (either
spectroscopic or photometric). Information on the core components were
retrieved from multifrequency, high resolution, high sensitivity
observations presented by
\citet{mo08,mo12,dd13,rossetti06,snellen00,cstan01,mo04,peck00,modd10}.\\
For the aforementioned sample, 
in Fig. \ref{peak} we plot the peak frequency (rest frame) versus the
largest linear size, $LLS$. 
The relation between the rest-frame peak frequency and
$LLS$, obtained by minimizing the chi-square error
statistic is: 

\begin{displaymath}
{\rm Log} \nu_{\rm p} = (-0.21\pm0.04) - (0.59\pm0.05) \times {\rm Log}
LLS
\end{displaymath}

\noindent which is in good agreement with the relation found by
\citet{odea98}. The HFP sources studied in this paper are in the
  top left region of Fig. \ref{peak}.
Interestingly, in the HFP sources departing
significantly from the relation, J0650+6001, J1335+5844, and
J1735+5049, the total spectrum
is dominated by a bright and very compact component,
like in J1335+5844 (see Fig. \ref{spectra}),
whose peak is above 5 GHz, while no significant contribution to the
spectrum from more extended features (mini-lobes) is present. 
The lack of diffuse emission in HFP sources 
may be a consequence of both the
high magnetic field (see Table \ref{magnetic}) which
causes severe radiative losses, and the short source lifetime which
prevents the formation of extended features.\\     

\subsubsection{Luminosity: core versus total}
\label{sec_core_tot}

\begin{figure}
\begin{center}
\includegraphics{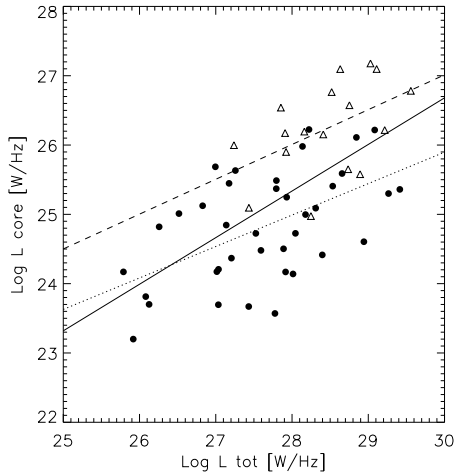}
\vspace{6.5cm}
\caption{Core luminosity at 5 GHz versus the source total luminosity
  at 375 MHz (rest frame) for the sources of the selected
  sample (Table \ref{samplone}). {\it Filled circles} refer to galaxies, while
  {\it empty triangles} are quasars. The solid
line represents the best linear fit for the whole sample, while the
dashed line and the dotted line represent the best fit considering
either quasars or galaxies, respectively 
(see Section \ref{sec_core_tot}). }
\label{lum-lum}
\end{center}
\end{figure}

To investigate the contribution of the core component to the total luminosity, 
in Fig. \ref{lum-lum} we plot the luminosity of the core at 5 GHz
versus the source total luminosity at 375 MHz (source rest frame). 
For the sources without
observations at 375 MHz, and 
for the GPS/HFP sources whose spectrum is absorbed at that frequency, the
375-MHz flux density was extrapolated from the optically thin part of
the spectrum. In the case the flux density of the core at 5 GHz was
not available, the 5-GHz flux density was extrapolated assuming a
flat spectral index ($\alpha = 0$).\\  
We performed a linear fit on the data displayed in Fig. \ref{lum-lum},
first considering the whole sample and subsequently separating
galaxies and quasars. If we consider the whole sample we obtain:

\begin{displaymath}
{\rm Log} L_{\rm core} = (6.51\pm3.17) + (0.67\pm0.11) \times {\rm Log}L_{\rm tot}
\end{displaymath}

\noindent which is similar to the relations derived 
for CSS/GPS objects \citep{dd13} and for FRI/FRII radio sources
\citep[e.g.][]{gg01,zirbel95,yuan12}. \\
If we consider the galaxies only we obtain:

\begin{displaymath}
{\rm Log} L_{\rm core, gal} = (12.28\pm3.26) + (0.45\pm0.12) \times
{\rm Log}L_{\rm tot, gal} 
\end{displaymath}

\noindent while for the quasars we find:

\begin{displaymath}
{\rm Log} L_{\rm core, q} = (11.94\pm6.64) + (0.50\pm0.23) \times {\rm
  Log}L_{\rm tot, q}
\end{displaymath}

\noindent Interestingly the slope derived for quasars (dashed line in
Fig. \ref{lum-lum}) is similar to the one derived for the galaxies
(dotted line in Fig. \ref{lum-lum}), and both are flatter than that of
the whole sample. However, there is a clear offset in the distribution
of $L_{\rm core}$ which is generally higher in quasars than in
galaxies. If we combine galaxies and quasars, we obtain an
artificially steepening of the slope that is likely due to the
dominance of the latter at high luminosities.\\

\begin{figure}
\begin{center}
\includegraphics{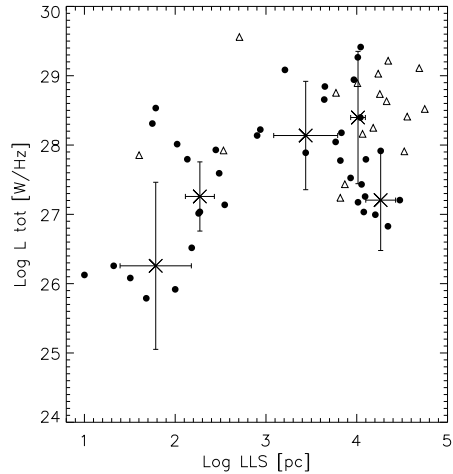}
\vspace{6.5cm}
\caption{Total luminosity at 375 MHz (rest frame) versus the
  largest linear size for the sources of the selected sample 
(Table \ref{samplone}). {\it Filled circles} refer to galaxies, 
while {\it empty
    triangles} are quasars. {\it Crosses} represent the median values
of the total luminosity and $LLS$, for galaxies only, separated in
different bins (see Section \ref{sec_tot_ls}). Error bars are the
standard deviations.}
\label{ptot-lls}
\end{center}
\end{figure}

\subsubsection{Total luminosity versus linear size}
\label{sec_tot_ls}

To determine how the luminosity evolves as the source grows,
in Fig. \ref{ptot-lls} we plot the total luminosity at 375
MHz (source rest frame), $L_{\rm tot}$, as a function of $LLS$. 
Quasars are mainly the most luminous and largest objects, 
while galaxies are found at
lower luminosity and smaller linear size. This arises from selection effects:
in pc-scale sources, the emission is dominated by
small components like the core or very compact hotspots.
As a consequence, the overall spectrum is flat and the source
  does not match the selection criteria of CSS/GPS samples.
Two outstanding examples are
the HFP sources J1335+5844 and J1735+5049 which were excluded from the
COINS sample of CSO candidates by \citet{peck00}
due to their flat overall spectrum, although
they turned out to be genuinely young radio sources (see Section 3.1).
On the other hand, when the
sources grow on larger scales, the contribution from the lobes dominates
making the overall spectrum steep enough to enter in the samples
selected on the basis of the spectral shape.\\
In order to prevent significant
contamination from selection and boosting effects, in the following
statistical analysis we consider only the galaxies. We divide the
galaxies
on the basis of the $LLS$ using five different bins containing
an equal number of objects. Median values are plotted as crosses and
the error bars represent their standard deviation. 
Fig. \ref{ptot-lls} suggests that the total radio luminosity
increases with $LLS$ up to about 1-10 kpc. As the source grows further
the total luminosity progressively decreases. These results are in
agreement with model predictions \citep[e.g.][]{snellen00b,fanti95},
in which the 
source expansion is influenced
by the surrounding ambient medium. The smaller sources are supposed to
reside within the innermost dense and inhomogeneous interstellar 
medium which may favour radiative losses. 
As the source grows, it experiences a less dense environment and the
adiabatic losses dominate, causing a decrease in the source luminosity.\\
The presence of large asymmetries, both in luminosity and arm-length,
mainly among the most compact objects \citep[][]{dd13,saikia03,mo08b},
indicates that during its growth one of the jets may interact with a
dense cloud. In this 
case its expansion is highly decelerated, while its luminosity is enhanced due to
severe radiative losses, making these objects common in flux-limited
samples. 
The asymmetric flux density
distribution in J1335+5844 and J1735+5049 and their very compact
hotspots may be the result of a strong interaction between a newly born
radio source and an inhomogeneous medium.\\

\subsubsection{Core luminosity evolution versus linear size}
\label{sec_core_ls}

\begin{figure}
\begin{center}
\includegraphics{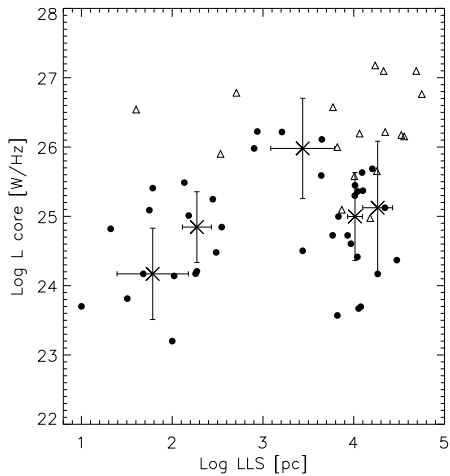}
\vspace{6.5cm}
\caption{Core luminosity versus the
  largest linear size for the sources of the selected sample (Table
  \ref{samplone}).  
{\it Filled circles} refer to galaxies, while {\it empty
    triangles} are quasars. {\it Crosses} represent the median values
of the core luminosity and $LLS$, for galaxies only, separated in
different bins (see Section \ref{sec_core_ls}). Error bars are the
standard deviations.}
\label{core-lls}
\end{center}
\end{figure}

In Fig. \ref{core-lls} we plot the core luminosity versus the source
linear size. The core luminosity spans roughly the same range of values
for both pc-scale and kpc-scale sources, 
although small objects have, on average, lower
core luminosity, while quasars are mostly in the upper right
region, likely due to boosting effects (see Section \ref{sec_core_tot}, and
Fig. \ref{lum-lum}).  \\ 
To investigate how the core luminosity evolves as the source grows, we
repeat the same statistical analysis performed for studying the
evolution of the source total luminosity. As in Section \ref{sec_tot_ls}
we consider only the galaxies and
we divide the data points in five bins on the
basis of their $LLS$, each bin containing
an equal number of objects. Crosses represent the median values and
the error bars represent the standard deviation.\\ 
The median values suggest an increase of the core luminosity with the
size for sources up to a few kpc, which is similar to what is found for the
total luminosity, although the poor statistics does not allow us to
confirm the peak around 1-10 kpc. \\
Interestingly, for the large sources ($LLS >$10 kpc) 
the core luminosity seems unrelated to the linear size, indicating a
different evolution for the core and total luminosity. The different
behaviour may be related to 
the influence of the surrounding medium which is supposed
at the basis of the total source luminosity evolution, but it should
not affect significantly 
the core luminosity. However, a larger number of objects
should be considered to confirm this result. Furthermore, we must keep
in mind that the core luminosity highly depends on the resolution of the
observations. If the resolution is not adequate the core emission may
be highly contaminated by the jets, causing an overestimate of its
luminosity. \\

\subsubsection{Core dominance versus linear size}
\label{sec_ratio_ls}

\begin{figure}
\begin{center}
\includegraphics{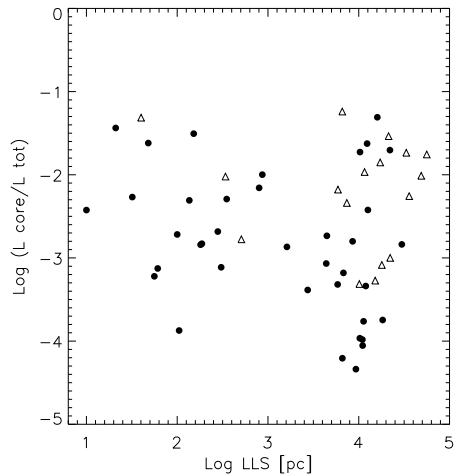}
\vspace{6.5cm}
\caption{Core luminosity-source total luminosity ratio versus the
  largest linear size for the sources of the
selected sample (Table \ref{samplone}). 
{\it Filled circles} refer to galaxies, while {\it empty
    triangles} are quasars.}
\label{ratio-lls}
\end{center}
\end{figure}

To investigate whether
the prominence of the core luminosity depends on the source size,
we plot the ratio of the core luminosity to the total luminosity, $L_{\rm
  core}/L_{\rm tot}$, versus the source size $LLS$ (Fig. \ref{ratio-lls}). No
clear trend is found between $L_{\rm
  core}/L_{\rm tot}$ and $LLS$, and $L_{\rm
  core}/L_{\rm tot}$ spans a large range of values both for pc-scale
and kpc-scale objects. Among the HFP sources, those with
the highest ratio are J0650+6001 and J0951+3451 whose core emission is
contaminated by the presence of additional contribution of the jet
that cannot be resolved even with VLBI observations. The HFP source with the
smallest ratio is J1335+5844 which is also the largest object among
the HFPs considered in this paper, and the
contribution from its hotspots is very dominant. Again, the largest
sources explore a wider range of ratios as a consequence of an
increased contribution of the extended emission, which is continuously
built as the source expands and ages. For the same reason, the absence of
objects in the bottom left part of Fig. \ref{ratio-lls} is likely due to 
the lack of significant extended features in the smallest sources, making
the core contribution a relatively high fraction of the
source total luminosity. \\

\section{Summary}

We presented multifrequency VLBA observations of four high frequency
peaking radio sources. The sources are characterized by a two-sided
structure and the radio emission is dominated by hotspots/lobes. The
core region is unambiguously detected in J1335+5844 and J1735+5049,
while in J0037+0808 we suggest that the central region is the
core, 
although observations at higher frequencies are needed to
confirm its nature. \\
The availability of multifrequency observations
covering both the optically-thick and -thin part of spectrum allowed
us to derive the peak parameters for the main source
components. The spectral peaks are above a few GHz, reaching
$\sim$10 GHz in the most compact hotspot of J1335+5844. 
Assuming that the spectral peak is produced by synchrotron
self-absorption we computed the magnetic field by means of observable
quantities. For half of the source components these values of the
magnetic field 
agree within a factor of a few, with the equipartition magnetic
field. However, in four components 
the magnetic field derived from observable quantities
is more than an order of magnitude higher than the equipartition
one. Such a discrepancy may be either caused by an inhomogeneous
component, not properly resolved by the observations presented here,
or it may indicate the presence of free-free absorption that modify
the optically-thick part of the spectrum, as it was observed in
other HFP galaxies \citep{mo08}, or it may witness a situation where
the radio emission is far from being in minimum energy condition.\\
To investigate possible changes in the physical properties as the
source grows, we built a sample of {\it bona-fide} young radio sources
spanning linear sizes from a few parsec to tens kpc. The sources
presented in this paper sample the small linear size tail of the distribution.\\
We confirm the
empirical correlation between the peak frequency and the linear
size and the relation between the core luminosity and the source total
luminosity. If we consider the source total luminosity as a function
of the linear size we find two different relations depending on the
source size: sources smaller than a few kpc increase their luminosity
as they grow, while larger sources progressively decrease the
luminosity, in agreement with the evolutionary models. The smallest
sources reside within the innermost region of the host galaxy where
the dense and inhomogeneous ambient medium favours radiative
losses. As the radio source expands on kpc scale it experiences a
smoother and less dense ambient medium and the adiabatic
losses dominate. The asymmetric radio structures of J1335+5844 and
J1735+5049 provide an additional support to the idea that the ambient
medium may influence the source growth even during the first
evolutionary stages.  \\
A different behaviour for sources smaller or larger than a few kpc is
also found for the core luminosity evolution. The former progressively
increase their core luminosity, reaching a peak at about a few kpc,
while for larger sources the luminosity seems unrelated with the
size. Such a trend may be partially contaminated by the inhomogeneous
resolution of the observations which depends on the source linear
size. For the smallest sources the core properties are measured on
milliarcsecond scale, while for the 
largest sources the core luminosity is derived on a region unresolved on 
arcsecond scale, and it may contain some contribution from the jet
base. \\

\section*{Acknowledgments}

We thank the reviewer A. Tzioumis for carefully reading the
manuscript and for giving valuable suggestions. 
The VLBA is operated by the US National Radio Astronomy Observatory which is a facility of the National Science Foundation operated under a cooperative agreement by Associated University, Inc., under contract with the National Science
Foundation. This research has made use of the NASA/IPAC Extragalactic
Database (NED), which is operated by the Jet Propulsion Laboratory,
California Institute of Technology, under contract with the National
Aeronautics and Space Administration.
Funding for SDSS-III has been provided by the Alfred P. Sloan Foundation, the Participating Institutions, the National Science Foundation, and the U.S. Department of Energy Office of Science. The SDSS-III web site is http://www.sdss3.org/.

SDSS-III is managed by the Astrophysical Research Consortium for the Participating Institutions of the SDSS-III Collaboration including the University of Arizona, the Brazilian Participation Group, Brookhaven National Laboratory, University of Cambridge, Carnegie Mellon University, University of Florida, the French Participation Group, the German Participation Group, Harvard University, the Instituto de Astrofisica de Canarias, the Michigan State/Notre Dame/JINA Participation Group, Johns Hopkins University, Lawrence Berkeley National Laboratory, Max Planck Institute for Astrophysics, Max Planck Institute for Extraterrestrial Physics, New Mexico State University, New York University, Ohio State University, Pennsylvania State University, University of Portsmouth, Princeton University, the Spanish Participation Group, University of Tokyo, University of Utah, Vanderbilt University, University of Virginia, University of Washington, and Yale University. 

\begin{table*}
\caption{The sample of {\it bona-fide} young radio sources. Column 1:
  source name; Column 2: redshift, p=photometric redshift; Column 3:
  largest linear size; Column 4: core luminosity at 5 GHz; Column 5:
  total luminosity at 375 MHz; Column 6: rest-frame peak frequency;
  Column 6: reference for the core component. a=this paper;
  b=\citet{owsianik98}; c=\citet{modd10}; d=\citet{mo08}; e=\citet{mo12};
  f=\citet{dd13}; g=\citet{murgia03}; h=\citet{peck00};
  i=\citet{rossetti06}; j=\citet{mo04}; k=\citet{cstan98};
  l=\citet{snellen00b}; m=\citet{rfanti90}.}
\begin{center}
\begin{tabular}{ccccccc}
\hline
Source&z&$LLS$&Log $L_{\rm core}$&Log $L_{\rm tot}$&$\nu_{\rm
  p}$&Ref\\
 & &kpc&[W/Hz]&[W/Hz]&GHz& \\
\hline
\multicolumn{7}{c}{From the sample presented by Dallacasa et al. (2000)}\\
\hline
J0111+3906 &  0.688 &   0.056 &   25.089 &      28.151 &  7.9 & b\\  
J0650+6001 &  0.455 &   0.040 &   26.54  &      27.74  &   8.0& c\\ 
J1335+5844 &  0.58p  &   0.105 &   24.140 &      27.874 &  8.8 &a\\ 
J1511+0518 &  0.084 &   0.010 &   23.701 &      26.101 &  10.9& d\\ 
J1735+5049 &  0.835 &   0.061 &   25.407 &      28.349 &   8.2& a\\
\hline
\multicolumn{7}{c}{From the sample selected by Stanghellini et al. (2009)}\\
\hline
J0951+3451 &  0.29p &   0.021 &   24.82  &      26.18  &   7.7& e\\
\hline
\multicolumn{7}{c}{From the sample selected by Dallacasa et al. (1995)}\\
\hline
0316+161   & 0.907  &   4.384 &   25.589 &      28.460&    1.5& f\\ 
0404+768   & 0.5985 &   0.866 &   26.224 &      28.080&    0.55&f\\ 
0428+205   & 0.219  &   0.351 &   24.845 &      27.077&    1.4 &f\\
1225+368   & 1.973  &   0.509 &   26.782 &      29.227&    5.2 &f\\
1323+321   & 0.369  &   0.305 &   24.480 &      27.497&    0.68&g\\ 
1358+625   & 0.431  &   0.280 &   25.247 &      27.821&    0.79&f\\ 
\hline
\multicolumn{7}{c}{From the sample selected by Taylor et al. (1996)}\\
\hline
0710+439   & 0.518  &   0.136 &   25.487 &      27.668&    2.88&h\\
2352+495   & 0.2379 &   0.187 &   24.207 &      26.971&    0.87&h\\
\hline
\multicolumn{7}{c}{From the sample selected by Fanti et al. (2001)}\\
\hline
0039+391 &   1.006 &    2.738 &   24.503 &      27.677&  $<$0.2&i\\
0120+405 &    0.84 &   18.345 &   24.169 &      27.730&  $<$0.18&i\\   
0137+401 &    1.62 &   35.939 &   26.153 &      28.117&  $<$0.26&1\\   
0213+412 &    0.515&   12.358 &   25.632 &      27.131&  0.20&i\\
0701+392 &    1.283&   15.167 &   24.975 &      27.997&  0.34&i\\
0744+464 &    2.926&   11.040 &   25.360 &      28.998&  $<$0.39&i\\
0814+441 &    0.12p&    8.556 &   23.057 &      25.547&  $<$0.112&i\\ 
0935+428a&    1.291&   10.969 &   24.415 &      28.145&  $<$0.46&i\\
0955+390 &    0.52p&   29.813 &   24.368 &      27.078&  $<$0.15&i\\  
1025+390b&    0.361&   16.022 &   25.686 &      26.901&  $<$0.14&i\\
1027+392 &    0.56 &   10.323 &   25.447 &      27.039&  0.23   &i\\
1157+460 &   0.7428&    5.852 &   24.726 &      27.876&  0.44   &i\\
1201+394 &   0.4449&   11.951 &   23.696 &      26.922&  $<$0.14&i\\ 
1242+410 &   0.813 &    0.340 &   25.900 &      27.741&    0.65&j\\
1458+433 &   0.927 &   12.608 &   25.370 &      27.594&  $<$0.19&i\\
2311+469 &   0.745 &   11.575 &   26.194 &      27.992&  $<$0.17&i\\
2349+410 &   2.046 &   10.146 &   25.579 &      28.554&  $<$0.30&i\\
2358+406 &   0.978 &    0.799 &   25.980 &      27.930&    0.59&j\\
\hline
\multicolumn{7}{c}{From the sample selected by Stanghellini et al. (1998)}\\
\hline
1345+125 &   0.122 &    0.152 &   25.011 &      26.482&   0.67&k\\
1943+546 &   0.263 &    0.181 &   24.173 &      26.943&   0.75&g\\
\hline
\multicolumn{7}{c}{From the sample selected by Snellen et al. (2001)}\\
\hline
1819+6707&   0.22  &    0.112 &   23.813 &      26.021&   0.976&l\\
1946+7048&   0.10  &    0.087 &   24.170 &      25.760&   1.98 &l\\
\hline
\end{tabular}
\end{center}
\label{samplone}
\end{table*}

\addtocounter{table}{-1} 
\begin{table*}
\caption{Continued.}
\begin{center}
\begin{tabular}{ccccccc}
\hline
Source&z&$LLS$&Log $L_{\rm core}$&Log $L_{\rm tot}$&$\nu_{\rm
  p}$&Ref\\
 & &kpc&[W/Hz]&[W/Hz]&GHz& \\
\hline
\multicolumn{7}{c}{From the sample selected by Fanti et al. (1990)}\\
\hline
3C\,43     &  1.459  &    22.170&   26.216 &      28.942&   $<$0.05&m\\
3C\,48     &  0.367  &     6.579&   25.998 &      27.143&      0.109&m\\
3C\,49     &  0.621  &     6.781&   24.997 &      28.030&   0.194&m\\
3C\,119    &  1.023  &     1.617&   26.217 &      28.870&   0.303&m\\
3C\,138    &  0.759  &     5.900&   26.574 &      28.581&   0.176&m\\
3C\,147    &  0.545  &     4.454&   26.110 &      28.712&   0.231&m\\
3C\,186    &  1.067  &    17.959&   25.651 &      28.515&   0.082&m\\
3C\,190    &  1.1944 &    33.356&   26.173 &      27.670&   0.088&m\\
3C\,216    &  0.6702 &    56.120&   26.764 &      28.364&   0.066&m\\
3C\,237    &  0.877  &     9.301&   24.605 &      28.751&   0.094&m\\
3C\,241    &  1.617  &    10.268&   25.300 &      28.973&   0.104&m\\
3C\,277.1  & 0.31978 &     7.392&   25.095 &      27.350&   $<$0.132&m\\ 
3C\,298    & 1.43732 &    21.297&   27.096 &      28.361&   0.195&m\\
3C\,309.1  &   0.905 &    17.215&   27.177 &      28.833&   $<$0.076&m\\
3C\,346    & 0.16201 &    22.056&   25.124 &      26.782&   $<$0.045&m\\
\hline
\end{tabular}
\end{center}
\end{table*}

\appendix
\section{VLA observations of bright HFP radio sources}

\subsection{VLA observations}

Multifrequency VLA observations of 13 of the 55 candidate HFPs from
the bright HFP sample \citep{dd00} were carried out on 2006 November
19 in C configuration, using
filler time. The observing bandwidth was chosen to be 50 MHz per
IF. Separate analysis for each IF in L, C and X bands was carried out
to improve the spectral coverage of the data, as it was done in previous
works \citep{dd00, tinti05, mo07, mo10}.
We obtained the flux density measurements in L band (IFs at 1.465 and
1.665 GHz), C band (4.565 and 4.935 GHz), X band (8.085 and 8.465
GHz), K band (22.460 GHz), and in Q band (43.340
GHz). Each source was typically observed for 50 s at each band,
cycling through frequencies. Therefore, the flux density measurements
can be considered almost simultaneous. 
About 3 min were spent on the primary flux density calibrator 3C 48,
while secondary calibrators, chosen to minimize the telescope slewing
time, were observed for 1.5 min at each frequency every $\sim$20 min. \\
The data reduction was carried out following the standard procedures
for the VLA, implemented in the \texttt{AIPS} package. 
In order to obtain accurate flux density
measurements in the L band, it was necessary to image several
confusing sources falling within the primary beam, and often
accounting for most of the flux density within the field of view. 
The final images were produced after a few phase-only self-calibration
iterations, and source parameters were measured by using the task
JMFIT, which performs a Gaussian fit on the image plane. 
The integrated flux density was
checked with TVSTAT. The flux density measurements at each frequency
and epoch are reported in Table \ref{vla-flux}. 
Apart from J0111+3908 which was already known to possess an
extended emission \citep{tinti05}, the other sources are unresolved with
the VLA in C configuration.\\ 
The uncertainty on the amplitude calibration, $\sigma_{\rm c}$, 
results to be within 3\%
in L, C and X band, and 10\% in K and Q bands. 
Strong RFI affected the 1.665 GHz data, causing larger amplitude
uncertainties of 
about 10\% at this frequency. \\
The rms noise level on the image
plane is not relevant for bright 
radio sources like our targets. In this case, the uncertainty on
the flux density, $\sigma_{\rm S}$, is roughly equal to 
$\sigma_{\rm c}$.\\ 

\begin{table*}
\caption{VLA flux density and spectral parameters. Col. 1: source
  name; Cols. 2, 3, 4, 5, 6, 7, 8, and 9: flux density at 1.4, 1.7,
  4.5, 4.9, 8.1, 8.4, 22, and 43 GHz; Cols. 10 and 11: peak frequency
  and peak flux density, respectively; Col. 12: variability index
  computed following \citet{mo07}.}
{\scriptsize
\begin{center}
\begin{tabular}{cccccccccccc}
\hline
Source&$S_{\rm 1.4}$&$S_{\rm 1.7}$&$S_{\rm 4.5}$&$S_{\rm 4.9}$&$S_{\rm
  8.1}$&$S_{\rm 8.4}$&$S_{\rm 22}$&$S_{\rm 43}$&$\nu_{p}$&$S_{p}$&V\\
  &mJy&mJy&mJy&mJy&mJy&mJy&mJy&mJy&GHz&mJy& \\
(1)&(2)&(3)&(4)&(5)&(6)&(7)&(8)&(9)&(10)&(11)&(12)\\
\hline
&&&&&&&&&&&\\
J0003+2129&  93$\pm$9& 113$\pm$17&  245$\pm$7&  247$\pm$7&  206$\pm$6&  199$\pm$6&  55$\pm$4&  14$\pm$2&5.0$\pm$0.3&241$\pm$2&13.6\\
J0005+0524& 182$\pm$18& 203$\pm$30&  203$\pm$6&  197$\pm$6&  170$\pm$5&  167$\pm$5&  93$\pm$6&  52$\pm$5&3.0$\pm$0.1&213$\pm$1&2.6\\
J0037+0808& 113$\pm$11& 125$\pm$19&  290$\pm$9&  297$\pm$9&  288$\pm$9&  285$\pm$9& 156$\pm$11&  83$\pm$8&7.0$\pm$0.1&290$\pm$1&2.0\\
J0111+3906& 437$\pm$44& 578$\pm$87& 1326$\pm$40& 1350$\pm$40&  969$\pm$30&  930$\pm$28& 290$\pm$20& 100$\pm$10&5.1$\pm$0.5&1244$\pm$4&0.4\\
J0116+2422& 104$\pm$10& 121$\pm$18&  242$\pm$7&  243$\pm$7&  234$\pm$7&  230$\pm$7& 135$\pm$13&  73$\pm$7&6.7$\pm$0.1&240$\pm$2&0.5\\
J0428+3259& 173$\pm$17& 208$\pm$31&  500$\pm$15&  510$\pm$15&  503$\pm$15&  497$\pm$15& 231$\pm$16&  94$\pm$9&6.5$\pm$0.1&516$\pm$2&2.0\\
J0625+4440& 127$\pm$13& 130$\pm$20&  144$\pm$4&  146$\pm$4&  146$\pm$4&  145$\pm$4& 122$\pm$8&  90$\pm$9&5.1$\pm$1.0&149$\pm$1&530.8\\
J0642+6758& 210$\pm$21& 257$\pm$39&  383$\pm$11&  381$\pm$11&  339$\pm$10&  332$\pm$10& 160$\pm$11&  71$\pm$7&4.9$\pm$0.2&378$\pm$1&7.4\\
J0646+4451& 746$\pm$75& 922$\pm$138& 2855$\pm$86& 2991$\pm$90& 3490$\pm$105& 3500$\pm$105&2676$\pm$187&1623$\pm$162&10.4$\pm$0.1&3524$\pm$1&7.7\\
J0650+6001& 569$\pm$57& 685$\pm$103& 1199$\pm$36& 1197$\pm$36& 1044$\pm$31& 1022$\pm$31& 523$\pm$37& 249$\pm$25&5.5$\pm$0.1&1146$\pm$3&1.4\\
J0655+4100& 242$\pm$24& 258$\pm$39&  363$\pm$11&  368$\pm$11&  369$\pm$11&  370$\pm$11& 333$\pm$23& 213$\pm$21&7.5$\pm$0.1&389$\pm$1&4.8\\
J2136+0041&4191$\pm$419&5256$\pm$788&10100$\pm$303&10079$\pm$302& 9111$\pm$273& 8847$\pm$265&4792$\pm$335&2749$\pm$275&6.2$\pm$0.1&9600$\pm$5&43.0\\
J2203+1007& 121$\pm$12& 156$\pm$23&  313$\pm$9&  311$\pm$9&  249$\pm$7&  239$\pm$7&  73$\pm$5&  24$\pm$2&4.9$\pm$0.2&300$\pm$1&0.4\\  
&&&&&&&&&&&\\
\hline
\end{tabular}
\end{center}}
\label{vla-flux}
\end{table*}

\subsection{Flux density variability}

The spectra obtained by the simultaneous multifrequency observations 
where fitted using the function discussed in
Section 3.2. The peak frequency and peak flux density are reported in
Table \ref{vla-flux}.\\
To investigate the presence of some variability, we compare the flux
densities with those measured in previous epochs and reported in
\citet{dd00,tinti05,mo07}. Following \citet{mo07} we computed the
variability index $V$ defined as:

\begin{displaymath}
V = \frac{1}{m} \sum^{m}_{i=1} \frac{(S_{i} - \bar{S_{i}})^{2}}{\sigma_{i}^{2}},
\end{displaymath}

\noindent where $S_{i}$ is the flux density at the $i$th frequency
measured at one epoch, $\overline S_{i}$ is the mean value of the flux
density computed by averaging the flux density at the $i$th frequency
measured at all the available epochs, $\sigma_{i}$ is the root mean
square (rms)  
on $S_{i} - \overline S_{i}$ and $m$ is the number of sampled frequencies. 
We prefer to compute the variability index for each epoch 
instead of considering all the epochs together in order to potentially
detect small outbursts.  \\
Significant variability is found in the sources J0625+4440 and
J2136+0041, which were already
classified as blazars and removed from the sample \citep{mo08a}. The
sources still considered young radio source candidates do not show
significant variability.\\

\end{document}